\documentclass[preprint,12pt,compress]{elsarticle}

\usepackage{setspace}
\usepackage{lineno}
\usepackage{graphicx,fancyhdr,amssymb}
\usepackage{color}
\usepackage{times,amsmath}
\usepackage[ps]{skak}
\usepackage{latexsym}
\usepackage{natbib}
\usepackage[numbers,sort&compress]{}
\usepackage{url,doi}
\usepackage{appendix}
\usepackage{hyperref}
\usepackage[linesnumbered,ruled]{algorithm2e}

\usepackage{longtable, booktabs} 
\usepackage{subfigure}
\usepackage{tabularray}
\usepackage{tabularx}
\usepackage{tcolorbox}
\usepackage{color} 
\usepackage{xcolor} 
\usepackage{pifont}
\usepackage{threeparttable}
\usepackage{multirow}
\usepackage{array}
\usepackage{makecell}
\usepackage{xurl}
\usepackage{longtable, xcolor, colortbl}
\usepackage{caption}
\usepackage{wasysym}
\usepackage{amssymb}
\usepackage{pifont}
\usepackage{tcolorbox}
\tcbuselibrary{breakable, skins}

    \newcounter{DaveCommentCounter}
\newcounter{DanyuCommentCounter}
\begin{document}

\begin{frontmatter}
\begin{nolinenumbers}


\title{\nolinenumbers Ribonucleic-Acid Protein Interaction Prediction Based on Deep Learning: A Comprehensive Survey}


\author[inst1]{Danyu Li}
\ead{3230002471@student.must.edu.mo}
\affiliation[inst1]{organization={School of Computer Science and Engineering, Macau University of Science and Technology},
           addressline={Taipa},
           city={Macau},
           postcode={999078},
           country={China}}

\author[inst1]{Rubing Huang\corref{cor1}}
\ead{rbhuang@must.edu.mo}
\cortext[cor1]{Corresponding author.}

\author[inst1]{Chenhui Cui}
\ead{3230002105@student.must.edu.mo}

\author[inst2]{Dave Towey}
\ead{dave.towey@nottingham.edu.cn}
\affiliation[inst2]{organization={School of Computer Science, University of Nottingham Ningbo China},
            addressline={Ningbo},
            city={Zhejiang},
            postcode={315100},
            country={China}}

\author[inst1]{Ling Zhou}
\ead{lzhou@must.edu.mo}

\author[inst1]{Jinyu Tian}
\ead{jytian@must.edu.mo}

\author[inst3]{Bin Zou}
\ead{binzou2009@ujs.edu.cn}
\affiliation[inst3]{organization={School of Food and Biological Engineering, Jiangsu University},
            addressline={Jiangsu},
            city={Zhenjiang},
            postcode={212013},
            country={China}}


\begin{abstract}
    The interaction between \textit{Ribonucleic Acids} (RNAs) and proteins, also called \textit{RNA Protein Interaction} (RPI), governs biological processes, including gene regulation and disease pathogenesis. 
    This comprehensive survey examines \textit{Artificial Intelligence} (AI) applications in \textit{Deep Learning-based RPI Prediction} (DL-based RPIP) through eight \textit{Research Questions} (RQs), analyzing 179 studies (2014--2023). 
    The key findings include: 
    sustained technical evolution through embryonic (2014--2017), accelerated (2018--2022), and expansion phases (2023) (RQ1);  
    hybrid models integrating \textit{Graph Neural Networks} (GNNs) (for topological interface modeling) and Transformers (for long-range dependencies) achieve state-of-the-art performance (RQ4); 
    pretrained language models enhance small-sample learning, but the cross-species generalization declines sharply with evolutionary distance (RQ5). 
    Critical challenges persist, including data heterogeneity across databases, the scarcity of standardized benchmarks (RQ2), and balancing the trade-off between feature encoding and information preservation (RQ3).
    Future advancements require biologically informed DL architectures, multi-feature fusion, and rigorous cross-validation to bridge the generalization-interpretability gap (RQ8):
    This would accelerate the clinical translation of predictive tools (RQ6/RQ7). 
    As the first comprehensive analysis spanning feature encoding, modeling, evaluation, applications, and tools, this work fills a critical gap in the DL-based RPIP literature.
\end{abstract}

\end{nolinenumbers}
\end{frontmatter}

\section{Introduction}

Ribonucleic Acids (RNAs) and proteins are important parts of living organisms, critical for the life activities of cells, including gene transcription and translation, and the regulation of biological functions.
RNA Protein Interaction (RPI) is involved in many cellular functions, such as protein synthesis~\cite{tuorto2012rna}, \textit{Messenger RNA} (mRNA) assembly~\cite{buchan2014mrnp}, viral replication~\cite{nagy2012dependence}, cell development regulation~\cite{keniry2012h19}, and \textit{Non-coding RNA} (ncRNA) function annotation~\cite{yang2015unveiling}.  
Previous studies have revealed that dysregulated RPI may indicate various diseases, such as cancers~\cite{bao2023protein,kang2020rna,okholm2020transcriptome} and neurological disorders~\cite{faghihi2008expression}.

Generally speaking, RPI is examined through biological experiments~\cite{ray2017rnacompete,cook2017rnacompete,xu2019enhanced}.
For example, \textit{Cross-Linking and Immunoprecipitation} (CLIP) checks RPI through ultraviolet crosslinking, which promotes covalent bonding between RNAs and proteins~\cite{wang2009clip}.
RNA immuno-precipitation explores RPI using known RNA and protein complexes~\cite{gagliardi2016rip}.
Although biological experiments can collect accurate results, they are generally expensive and time-consuming, especially when processing large amounts of complex data:
CLIP, for example, may require several days or weeks~\cite{konig2010iclip}.
In addition, expensive equipment is also often required~\cite{torres2018rna}:
CLIP requires 100 million to 1 billion biological cells, which means that the time and cost involved in expanding RPI experiments can be overwhelming~\cite{ramanathan2019methods}.

RPI Prediction (RPIP) has been developed to save time and to reduce costs:
It predicts the RPI using computational methods, such as matrix factorization, and probabilistic approaches~\cite{zhang2018lpgnmf,pai2017sequence}.
As discussed by Pan et al.~\cite{pan2019recent}, RPIP consists of the following three tasks:

\begin{itemize}
    \item[(1)] Predicting whether or not there will be an interaction between the RNAs and proteins. 
    \item[(2)] Predicting the RPI binding sites (locations of interaction) or motifs (repeated patterns or structures). 
    \item[(3)] Predicting whether or not a protein is an \textit{RNA Binding Protein} (RBP).
\end{itemize}

Unfortunately, traditional computational methods can be slow, and may not be able to deal with large and complex biological data.
Artificial intelligence, especially \textit{Deep Learning} (DL), has demonstrated remarkable potential for processing massive amounts of data in various applications~\cite{sagar2019recent,sorin2020deep,jiao2019survey}.
DL can learn high-dimensional features from large amounts of complex RNA and protein data, and can provide accurate RPIP results.
DL-based RPIP is, therefore, an important research direction.

Over the last decade, DL-based RPIP has had many successes.
Although there have been some papers presenting overviews of RPIP, they neither covered the entire procedure nor focused on DL-based RPIP. 
The summary of DL-based RPIP by Pan et al.~\cite{pan2019recent}, for example, only included three types of DL models.
Similarly, Si et al.~\cite{si2015computational} reviewed studies of predictions of RNA-binding sites and RBPs using \textit{Machine Learning} (ML) models, not DL models.
To date, there has been no exhaustive survey summarizing the state-of-the-art of DL-based RPIP:
This paper aims to fill this gap in the literature.

In this paper, we provide a comprehensive survey of DL-based RPIP, covering 179 papers published between 2014 and 2023. 
This paper provides the following eight contributions:
(1) a classification and description of RPI datasets; 
(2) an introduction to RNA and protein feature-encoding methods; 
(3) a description, classification, and analysis of DL-based RPIP models; 
(4) a summary of model-performance evaluations and metrics; 
(5) a discussion of DL-based RPIP applications;
(6) a summary of available DL-based RPIP websites and software;  
(7) a discussion of some remaining DL-based RPIP research challenges and potential future work; and 
(8) a summary and analysis of the relevant literature.
To the best of our knowledge, this is the most comprehensive large-scale survey of DL-based RPIP to date.

The rest of this paper is organized as follows. 
Section~\ref{SEC:Background} introduces the prerequisite knowledge of RPIP.
Section~\ref{SEC:Methodology} discusses the literature review's methodology. 
Section~\ref{SEC:rq1} examines the evolution and distribution of DL-based RPIP. 
Section~\ref{SEC:rq2} presents existing RPIP datasets. 
Section~\ref{SEC:rq3} discusses the feature extraction and encoding of RNAs and proteins.
Section~\ref{SEC:rq4} presents the DL models used for RPIP. 
Section~\ref{SEC:rq5} gives a detailed analysis of RPIP evaluation metrics. 
Section~\ref{SEC:rq6} summarizes some applications of RPIP. 
Section~\ref{SEC:rq7} introduces some websites and software for RPIP. 
Section~\ref{SEC:rq8} analyzes some of the potential challenges to be addressed in future work.
Section~\ref{SEC:Related_Survey} presents a comparison of this paper with other review papers, highlighting the innovations in our study.
Finally, Section~\ref{SEC:Conclusion} concludes the paper.

\begin{figure}[!t]
    \centering
    \includegraphics[width=\textwidth]{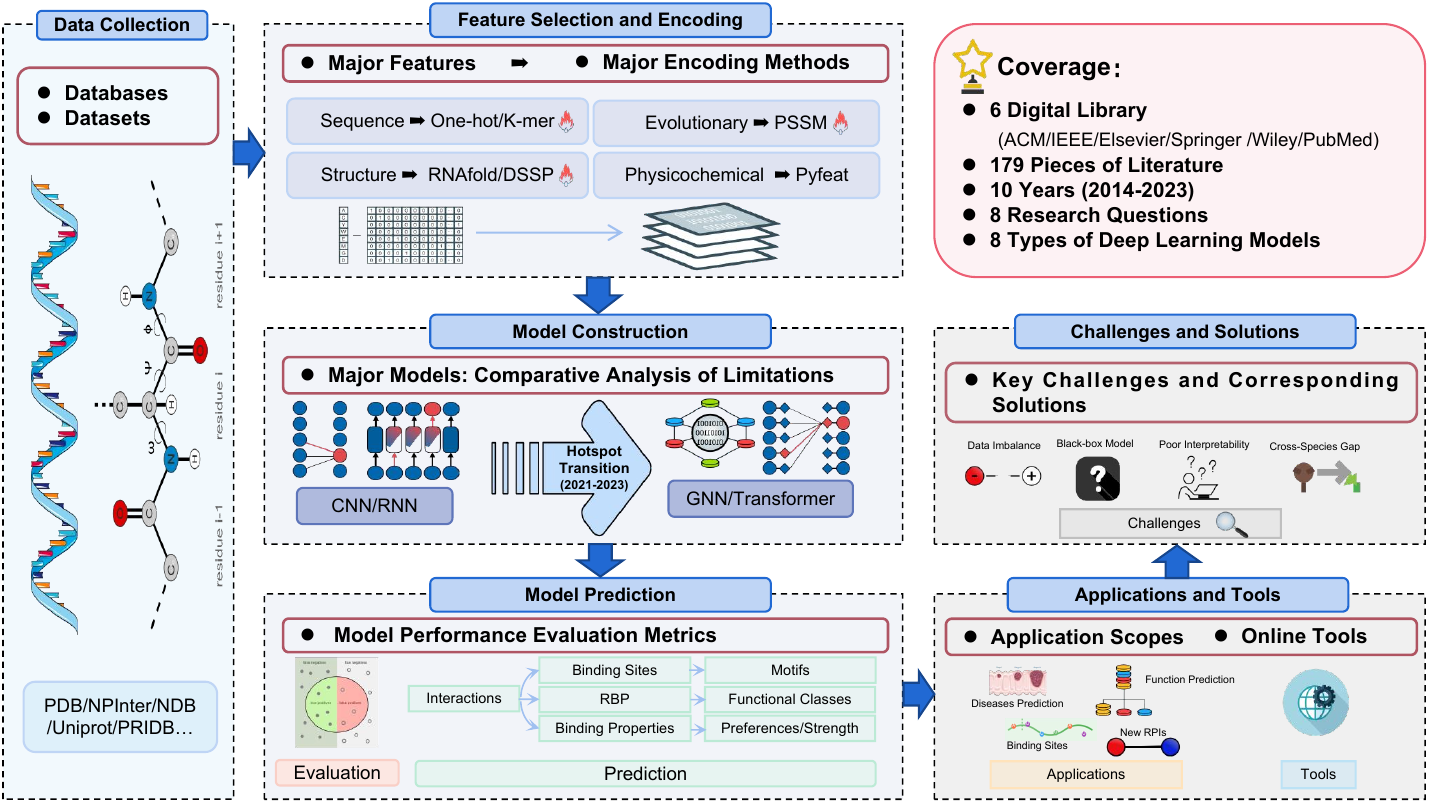}
    \caption{DL-based RPIP graphical abstract.}
    \label{FIG: Graphical_abstract}
\end{figure}

Figure~\ref{FIG: Graphical_abstract} presents the entire workflow of DL-based RPIP.
The workflow includes the following core modules:
(1) {\em Data Collection} integrates ten databases and 11 datasets, covering 179 literature items, aiming to answer eight key research questions.
(2) {\em Feature Processing} selects five feature types and, for these features, discusses six major categories of encoding methods.
(3) {\em Model Construction}  compares eight types of DL models, with a focus on analyzing the technological evolution and limitations.
(4) {\em Prediction and Evaluation} uses eight performance evaluation metrics to assess interactions, binding sites, and RBP functions, covering the verification of prediction results.
(5) {\em Challenges and Solutions} proposes potential solutions for six major challenges (database fragmentation, dataset bias, feature encoding limitations, cross-species generalization gaps, model validation gap, and deployment opacity).
(6) {\em Applications and Tools} 
combines seven major application scenarios and 20 online tools to promote the implementation of research.
The overall design forms a closed-loop framework:
data $\rightarrow$ features $\rightarrow$ models $\rightarrow$ evaluation $\rightarrow$ applications, emphasizing the integrated research of AI and biomedicine.

\begingroup 
\begin{longtable}
{>{\raggedright\arraybackslash \scriptsize}m{2cm}@{\hspace{6pt}}
>{\raggedright\arraybackslash \scriptsize}m{4cm} !{\vline}
>{\raggedright\arraybackslash \scriptsize}m{2cm}@{\hspace{6pt}} 
>{\raggedright\arraybackslash \scriptsize}m{4cm} } 
\captionsetup{font=footnotesize}
\caption{{List of abbreviations.}}
\setlength{\tabcolsep}{1pt} 
\label{tab:abbreviations} 
\\\hline
\textbf{Abbreviation} & \textbf{Full name} & \textbf{Abbreviation} & \textbf{Full name} \\\hline 
\endfirsthead
\caption[]{{Continued.}} \\\hline
\textbf{Abbreviation} & \textbf{Full name} & \textbf{Abbreviation} & \textbf{Full name} \\\hline 
\endhead
\hline 
\multicolumn{4}{r}{{\scriptsize {Continued on next page.}}} \\ 
\endfoot
\hline 
\endlastfoot 
3D	& Three-Dimensional	& LSTM	& Long Short-Term Memory
\\\hline
A	& Angstroms & MCC &	Matthews Correlation Coefficient
\\\hline
AI	& Artificial Intelligence	& MEME	& Multiple Expectation Maximization for Motif Elicitation					
\\\hline
API &	Application Programming Interface&	miRNA	& MicroRNA	
\\\hline
ATH	& Arabidopsis Thaliana	& ML& Machine Learning	
\\\hline
AUPR&	Recall-Precision Curve Area	& MLP	& Multi-Layer Perceptron 
\\\hline
AUROC &	Area Under the ROC Curve&	mRNA	& Messenger RNA
\\\hline
BERT	&Bidirectional Encoder Representation from Transformers 	& MSE	& Mean Square Error										
\\\hline
BiLSTM	& Bidirectional Long Short-Term Memory	& NABind& Nucleic Acid Binding Predictor 									
\\\hline
BioGRID	& Biological General Repository for Interaction Datasets	& ncRNA	& Non-coding RNA										
\\\hline
BioSeq2vec	& Biological Sequence to Vector &	NDB	& Nucleic Acid Database								
\\\hline
BioTriangle &	 Biomolecule Tri-class Angle Platform for Integrated Exploration	&NONCODE	&Integrated Knowledge Database of Non-coding RNAs									
\\\hline
CircBase&	Database for Circular RNAs &	NPInter	&Non-coding RNA and Protein Interaction Database					
\\\hline
CircInteractome	& CircRNA Interactome Database	& NR	&Not Reported		
\\\hline
CircRNA	& Circular RNA	& NucleicNet	& Nucleic Acid-Binding Neural Network	
\\\hline
circRNA2Vec	& circRNA to Vectors&	PDB	&Protein Data Bank
\\\hline
CircRNADb &	Circular RNA Database&	PlncRNADB	&Plant Long noncoding RNA Database 										
\\\hline
CLIP &	Cross-Linking and Immunoprecipitation 	&PNG &	Portable Network Graphics									
\\\hline
CLIPdb	& CLIP-sequencing Database &POSTAR	&POST-trAnscriptional Regulation 										
\\\hline
CNNs &	Convolutional Neural Networks	&POV& Persistence of Vision Raytracer										
\\\hline
CRWS&	CircRNA-RBP Web Server&	Pprint2	& Protein-RNA Interface Predictor version 2					
\\\hline
CSV	& Comma-Separated Values	& PredDRBP-MLP& 	Predictor of DNA-Binding Proteins and RNA-Binding Proteins - Multilayer Perceptron	
\\\hline
DAE	& Denoising Autoencoder	& PRIDB	& Protein-RNA Interface Database 
\\\hline
DBNs&	Deep Belief Networks &	PrismNet &	Protein-RNA Interaction by Structure-informed Modeling using deep neural NETwork 
\\\hline
DeepCLIP& Deep Contextual Learning for Identifying Protein-RNA Interaction Profiles	& Pro2vec& 	Protein to Vectors	
\\\hline
DeepBind &	Deep Learning Model for Binding Site Specificity Prediction	& ProNA2020	& Protein--Nucleic Acid 2020
\\\hline
DeepDISOBind &	Deep Disorder Interaction Specificity for Oligonucleotides and Proteins Binding	& ProtT5-XL	& Protein Text-to-Text Transfer Transformer - Xtra Large				
\\\hline
DisProt	& Disordered Proteins Database	& ProtTrans	& Protein Transformer				
\\\hline
DL	& Deep Learning	& Pse-in-One &	Pseudo components in One-stop shop				
\\\hline
DL-based RPIP	& RPIP based on Deep Learning &	PSFM & Position Specific Frequency Matrix							
\\\hline
DLPRB &	Deep Learning for Protein-RNA Binding &	PSI-BLAST	& Position-Specific Iterated Basic Local Alignment Search Tool					
\\\hline
DNA	& Deoxyribonucleic Acid	& PSSM	& Position Specific Scoring Matrix						
\\\hline
DNABERT &	DeoxyriboNucleic-Acid BERT	& PST-PRNA	& Protein Surface Topography for Protein-RNA Binding Sites Prediction	
\\\hline
DNNs& Deep Neural Networks	& Pyfeat	&Python-based Effective Feature Generation Tool					
\\\hline
doc2vec &	Document to Vector 	& RBP	& RNA Binding Protein		
\\\hline
DoRiNA	&Database of RNA Interactions	&RBPDB	& RNA-Binding Protein Database					
\\\hline
DSSP&	Define Secondary Structure of Proteins &	RBPNet	& RNA Binding Protein Network					
\\\hline
ENCODE&	Encyclopedia of DNA Elements&	RBPsuite &	RNA-Binding Protein Prediction Suite			
\\\hline
ENCORI	& Encyclopedia of RNA Interactomes &	ResNet	& Residual Network					
\\\hline
ESM	& Evolutionary Scale Modeling	& RNA2vec	& RNA to Vectors 	
\\\hline
FN	& False Negatives &	RNACentral	& RNA Central Database
\\\hline
FP &	False Positives	& RNAFM	& RNA Foundation Model
\\\hline
FPR	& True Positive Rate	& RNAFormer	& RNA Fold Oriented Representation Model with Transformer		
\\\hline
GANs &	Generative Adversarial Networks	& RNAincoder &	RNA Interaction Coder					
\\\hline
GAT	& Graph Attention network	& RNAs& 	Ribonucleic Acids
\\\hline
GCN	 & Graph Convolutional Network	& RNNs&	Recurrent Neural Networks 
\\\hline
GEO &	Gene Expression Omnibus	& RPI	& RNA Protein Interaction 
\\\hline
GIF &	Graphics Interchange Format	& RPIP 	& RNA Protein Interaction  Prediction	
\\\hline
GNNs	& Graph Neural Networks &	RQs	& Research Questions
\\\hline
GraphBind &	Graph-based Binding residue Identification	&snoRNA	& Small Nucleolar RNA					
\\\hline
GraphSAGE	& Graph Sample and Aggregate&	SOPMA	& Self-Optimized Prediction Method with Alignment					
\\\hline
GRU	& Gated Recurrent Unit &	SPOT-RNA	& Single-sequence-based Prediction Of Tertiary RNA structures					
\\\hline
GTF	& Gene Transfer Format	& SSPro	& Protein Secondary Structure Predictor 					
\\\hline
HHblits&	Hierarchical Hidden Markov Model-based Blind  Iterative  Threading Searcher	&ThermoNet	&Thermodynamic Prediction		
\\\hline
HTML&	HyperText Markup Language	& TN	&True Negatives	
\\\hline
iDeepMV &	Integrated Deep Multi-View Learning Framework for RNA-Binding Protein Recognition&	TP	&True Positives					
\\\hline
iDRBP-ECHF	&Identifying DNA- and RNA-Binding Proteins based on Extensible Cubic Hybrid Framework	&TPR&	False Positive Rate	
\\\hline
iDRNA-ITF&	Inductive DNA-RNA Binding Residue Identification via Induction-Transfer Framework &	tRNA&	Transfer RNA
\\\hline
IPMiner	& Interaction Pattern Miner	& TSV&	Tab-Separated Values
\\\hline
JAR3D&	Java-based Alignment of RNA using 3-Dimensional Structure	&TXT&	Text File					
\\\hline
JPG	&Joint Photographic Experts Group	&Uniprot	&Universal Protein Resource					
\\\hline
JSON &	JavaScript Object Notation	& URL&	Uniform Resource Locator					
\\\hline
LncRNA	& Long non-coding RNA	& VGAE	& Variational Graph Auto-Eencoder
\\\hline
LncRNAdb	&Long non-coding RNAs Database	& word2vec	& Word to Vector	
\\\hline
LPI &	Long non-coding RNA Protein Interaction	& ZEA	& Zea Mays 
\\\hline
\end{longtable}
\endgroup

Table~\ref{tab:abbreviations} provides a list of abbreviations used in this paper.

\section{Background
\label{SEC:Background}}

This section introduces some of the preliminary concepts related to RNAs, proteins, RPI, and RPIP.
Some previous survey papers related to this topic are also reviewed.

\subsection{RNAs and proteins}
RNA sequences consist of four nucleotide bases: 
\textit{Adenine}; 
\textit{Cytosine};  
\textit{Guanine}; and 
\textit{Uracil}. 
The secondary structure of RNA is formed through the action of non-covalent bonds (mainly hydrogen bonds) between some nucleotide bases.
Folding of these secondary structures results in the \textit{three-Dimensional} (3D) RNA structures \cite{washietl2009sequence}.

Protein sequences are linear chains composed of 20 common amino acids. 
These linear structures undergo folding and coiling to form stable 3D structures, known as protein structures~\cite{zheng2014proteins}.
These sequences and structures contain detailed biochemical features, such as dipeptide composition~\cite{wang2021method} and atomic distances or angles~\cite{wang2022idrna}.

\subsection{RNA-protein interactions}

RPIs are essential for many cellular processes, including protein synthesis~\cite{noller2012evolution} and RNA transport~\cite{moffett1983characterization}. 
Research has also shown that proteins can interact with various types of RNA, such as mRNAs~\cite{prado2018cellulo}, \textit{Long Non-coding RNAs} (LncRNAs)~\cite{ferre2016revealing}, and \textit{circular RNAs} (circRNAs)~\cite{du2017identifying}.

Many methods have been developed to identify RPIs.
One commonly-used RPI approach involves calculation of the least atom distances between the RNAs and proteins~\cite{si2015computational}, typically measured in \textit{Angstroms} (\AA).
An RPI is generally considered to occur when the distance is less than a specific threshold, commonly set at values like 3.4~\AA~\cite{allers2001structure}.
A least atom distance of less than 3.4~\AA~usually indicates significant interactions.

\subsection{RPI prediction}

RPIP predicts interactions using RNA and protein information.
Recent RPIP research can be broadly categorized into three main types: 
\textit{biological experiment-based RPIP}; 
\textit{ML-based RPIP}; and 
\textit{DL-based RPIP}.

\subsubsection{Biological-experiment-based RPIP}
Various biological experiments have been used in RPIP.
RNA-based RPIP methods include electrophoretic mobility shift assay~\cite{bak2015electrophoretic}, chromatin isolation by RNA purification~\cite{chu2012chromatin}, and RNA pull-down protocol~\cite{torres2018rna}.
Protein-based approaches include  RNA immunoprecipitation~\cite{gagliardi2016rip} and cross-linking and immunoprecipitation~\cite{wang2009clip}.
Although biological experiments can produce highly accurate results, and have facilitated the construction of many RNA and protein datasets (which are valuable resources for RPIP), these experiments are often time-consuming and can require substantial human and experimental resources~\cite{ramanathan2019methods}.

\subsubsection{ML-based RPIP}
ML has been successfully applied to RPIP~\cite{mishra2021airbp,hossain2023pcpi}.
Supervised learning (including \textit{Support Vector Machines}~\cite{kumar2008prediction}, \textit{Random Forests}~\cite{pan2014predicting}, \textit{Naive Bayes}~\cite{hong2023random}, and other classifiers~\cite{jain2018data, mishra2021airbp}) has been used to capture the relationships between RNA-protein-features and labels.
In contrast, unsupervised learning focuses on identifying similarities within the data to make predictions, even when only unlabelled data is available~\cite{mamitsuka2005essential,theofilatos2015predicting}.
A key factor in the predictive performance of ML models is the effective extraction of features from RNAs and proteins. 
However, this process often requires manual feature extraction. 
Due to the diversity and complexity of RNA and protein features, manually extracting features can lead to reduced effectiveness.

\subsubsection{DL-based RPIP}
\textit{Deep Neural Networks} (DNNs) have revolutionized the field of biology by not only optimizing various feature representations, but also integrating the knowledge acquired during the learning process. 
DL can model highly complex structures, such as the sequences and structures of RNAs and proteins. 
Different DL models have evolved for various RPIP processes, including:
CNNs~\cite{shibu2023prediction}; 
RNNs~\cite{wang2021prediction};
\textit{Generative Adversarial Networks} (GANs)~\cite{wang2023rpi};
\textit{Deep Belief Networks} (DBNs)~\cite{du2018deepmvf};
\textit{Denoising Autoencoder} (DAE)~\cite{pan2016ipminer};
\textit{Multi-Layer Perceptron} (MLP)~\cite{arican2023preddrbp};
GNNs~\cite{wei2023lpih2v}; 
and \textit{Transformers}~\cite{yamada2022prediction}. 
This paper analyzes DL-based RPIP approaches in detail in Section~\ref{SEC:rq4}.

\section{Methodology
\label{SEC:Methodology}}

\begin{figure}
    \centering
    \includegraphics[width=\textwidth]{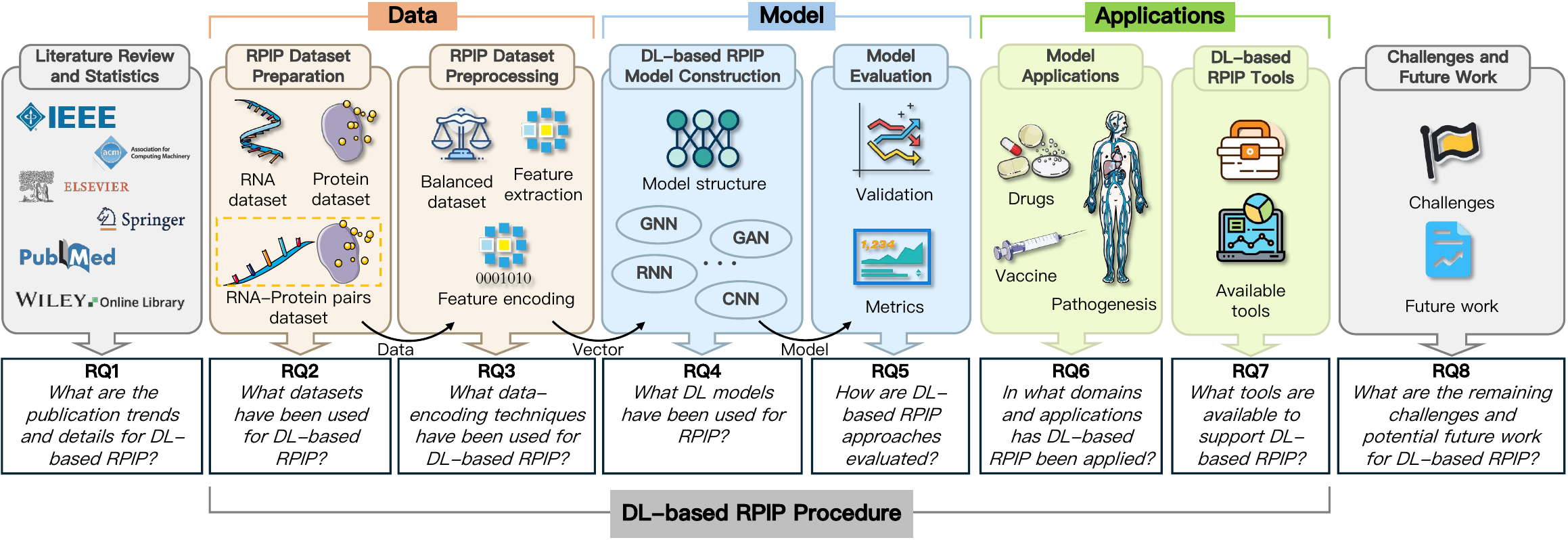}
    \caption{The framework of this survey.}
    \label{FIG:Framework}
\end{figure}

In this paper, we followed a structured and systematic method to perform the DL-based RPIP survey. 
The detailed methodology used is described in this section.
The research process of the survey is shown in Figure \ref{FIG:Framework}.

\subsection{Research questions}
The goal of this survey is to summarize the state of research and challenges for DL-based RPIP.
This paper classifies these things based on details and evidence from data encoding for DL-model construction.
To achieve this, we were guided by the following research questions (RQs):
\begin{itemize}
    \item {\textbf{RQ1}}: What are the publication trends and details for DL-based RPIP?
    \item {\textbf{RQ2}}: What datasets have been used for DL-based RPIP?
    \item {\textbf{RQ3}}: What data-encoding techniques have been used for DL-based RPIP?
    \item {\textbf{RQ4}}: What DL models have been used for RPIP?
    \item {\textbf{RQ5}}: How are DL-based RPIP approaches evaluated?
    \item {\textbf{RQ6}}: In what domains and applications has DL-based RPIP been applied?
    \item {\textbf{RQ7}}: What tools are available to support DL-based RPIP?
    \item {\textbf{RQ8}}: What are the remaining challenges and potential future work for DL-based RPIP?
\end{itemize}

The answer to RQ1 will provide an overview of the published DL-based RPIP work.
The answer to RQ2 will identify existing datasets for DL-based RPIP.
The answer to RQ3 will provide commonly used encoding methods for RNA and protein features.
The answer to RQ4 will explain currently-used DL models for RPIP.
The answer to RQ5 will explore how to evaluate RPIP model performance, including both evaluation experiments and metrics.
The answer to RQ6 will list where and how DL-based RPIP has been applied.
The answer to RQ7 will explain what RPIP tools have been developed.
Finally, the answer to RQ8 will provide an overview of some remaining challenges and potential research opportunities in RPIP.

\subsection{Literature search and selection}
The literature review used the following six online literature repositories belonging to publishers of technical research~\cite{huang2019survey}:

\begin{itemize}
    \item ACM Digital Library (https://dl.acm.org)
    \item IEEE Xplore Digital Library (https://ieeexplore.ieee.org/Xplore/home.jsp)
    \item Elsevier Science Direct (https://www.sciencedirect.com)
    \item Springer Online Library (https://link.springer.com)
    \item Wiley Online Library (https://onlinelibrary.wiley.com)
    \item PubMed (https://pubmed.ncbi.nlm.nih.gov/?db=PubMed)
\end{itemize}

The choice of these repositories was influenced by the fact that many important journal articles about DL-based RPIP are available through Elsevier Science Direct, Springer Online Library, and Wiley Online Library. 
Additionally, the ACM Digital Library and the IEEE Xplore Digital Library not only offer articles from conferences, but also provide access to some important relevant journals.
PubMed comprises biomedical literature from MEDLINE, life science journals, and online books:
It is currently the most widely used free online database for medical literature, and includes many articles on the topic of RPIP.

After the literature repositories were determined, each one was searched according to the survey parameters.
The queries were formulated around the topic of DL-based RPIP.
Each repository, except Springer, had support for conducting advanced searches. 
Because it appeared that an advanced search could not be formulated in Springer, a broad search query was instead conducted, and limited to our keywords:
The most relevant 400 articles were then selected. 
We used an advanced search in other repositories.
IEEE provides a command search, and PubMed has a query box, these functions were convenient for the search. 
ScienceDirect was limited to a maximum of eight boolean operators, so we chose as many keywords as possible in the search scope. 
Another challenge with ScienceDirect was that it
did not support wildcards.
The six repositories listed in Table~ \ref{tab:search query} were used to find relevant papers.

\begin{table}[!t]
  \caption{Selected repositories with search queries.}
  \scriptsize
  \centering
  \label{tab:search query}
  \setlength{\tabcolsep}{1mm}{
          \begin{tabular}{c|m{11.4cm}}\hline
  \textbf{Digital library}& \multicolumn{1}{c}{\textbf{Search query}} \\\hline                       
  ACM &  [[All: "rna protein"] OR [All: "rbp"] OR [All: "lpi"] OR [All: "rpi"]] AND [[All: "interaction"] OR [All: "binding"]] AND [[All: "prediction"] OR [All: "model"] OR [All: "learning"] OR [All: "network"]]
 \\\hline
  IEEE & (("Document Title":"$\ast$RNA" AND "Document Title":"protein$\ast$") OR "Document Title":"$\ast$RBP" OR "Document Title":"$\ast$RPI" OR "Document Title":"LPI" OR "Document Title":"$\ast$RNA Binding") AND  ("Document Title":"Interaction$\ast$" OR "Document Title":"Binding")  AND ("Document Title":"prediction" OR "Document Title":"model$\ast$" OR "Document Title":"deep" OR "Document Title":"graph"  OR "Document Title":"learning" OR "Document Title":"network$\ast$")
  \\\hline
  Elsevier & Title, abstract or author-specified keywords: (("RNA protein") OR ("RBP") OR ("RPI")) AND (("Interaction") OR ("Binding")) AND (("prediction") OR ("model") OR ("learning") OR ("network"))
 \\\hline
  Springer & ((protein AND RNA) OR (RBP) OR (LPI) OR (RPI)) AND  ((Interaction) OR (binding)) AND ((predict$\ast$) OR (model) OR (learning) OR (network) OR (deep) OR (graph))
 \\\hline
  Wiley & ""RNA protein" " anywhere and ""binding" "interaction"" anywhere and ""deep" "network" " learning" "model" "predict$\ast$"" anywhere
 \\\hline
  PubMed & ((protein[Title]) AND ($\ast$RNA[Title])) AND ((Interaction[Title]) OR (binding[Title])) AND ((predict$\ast$[Title]) OR (deep[Title]) OR (learning[Title]) OR (model$\ast$[Title]) OR (network$\ast$[Title]) OR (graph[Title]))
  \\\hline   
  \end{tabular}  
  }
\end{table}

\subsection{Paper selection criteria}
Table~\ref{tab:criteria} lists the inclusion criteria used to select papers. 
Generally, the chosen papers were written in English and related to DL-based RPIP.
We excluded these (e.g., M.Sc. and Ph.D.) and existing surveys (as discussed in Section ~\ref{SEC:Related_Survey}).
We also confined the literature to those articles that were freely available (e.g., through an open-access publication policy, or without fees).

\begin{figure}[!b]
\centering
\includegraphics[width=0.35\textwidth]{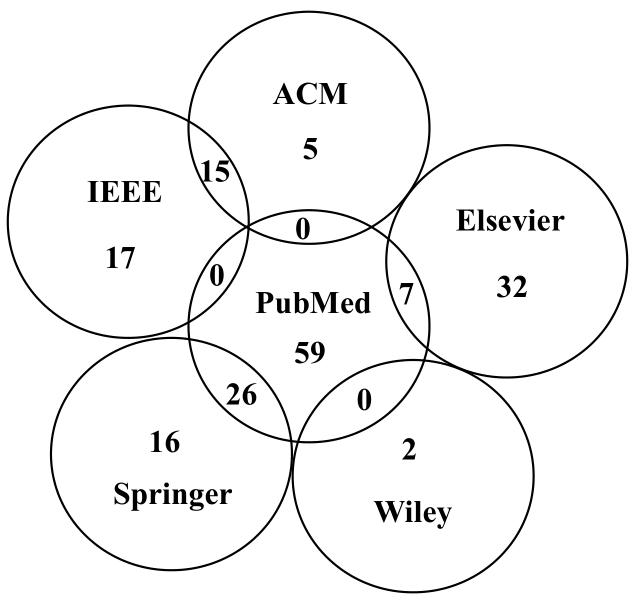}
\caption{Search engines and the number of primary studies.}
\label{FIG: search engines}
\end{figure}

Removal of duplicates and application of the exclusion
criteria reduced the initial 1897 candidate studies to 172 published papers. 
Finally, a snowballing process~\cite{golmohammadi2023testing} 
was conducted by checking the references of the selected 172 papers, resulting in the addition of seven more papers. 
In total, 179 publications (primary studies) were included in the survey.
The specific details are shown in Table~\ref{tab:selection_result}.
The number and duplication of papers filtered from each repository are shown in Figure~\ref{FIG: search engines}.

We acknowledge that it may be impossible to find all DL-based RPIP papers using only our search. 
However, we are confident that we have included most of the relevant published papers, and that our survey provides the overall trends and the state-of-the-art of DL-based RPIP.

\subsection{Data extraction and collection}

We carefully read and examined all 179 primary studies, extracting data based on the research questions. 
As shown in Table~\ref{tab:data extraction}, we identified the following information from each study: 
databases and datasets, features and encoding methods, DL models, model-evaluation methods and metrics, applications, tools (such as websites or software), and remaining challenges for RPIP.
The data extraction involved dividing the papers among the authors, with the extraction being performed by one
author and the results being double-checked by another.
In case of any discrepancy, the issue was discussed and settled by a third author.

\begin{table}[!t]
  \caption{Selection criteria.}
  \scriptsize
  \centering
  \label{tab:criteria}

  \setlength{\tabcolsep}{4mm}{
          \begin{tabular}{c|l}\hline
  \textbf{No.}& \textbf{Criterion} \\\hline                             
  1 & The paper is in English \\\hline
  2 & The paper is related to RPI  \\\hline
  3 & The paper is about DL and RPI (but does not have to be exclusively about DL and RPI) \\\hline
  4 & The paper is not a thesis \\\hline
  5 & The paper is not a survey or a systematic literature review \\\hline
  6 & The paper should be freely available (e.g., open access) \\\hline
  \end{tabular}  }     
\end{table}

\begin{table}[!t]
  \caption{Selection results of primary studies.}
  \centering
  \scriptsize
  \label{tab:selection_result}
 \setlength{\tabcolsep}{2.6mm}{
          \begin{tabular}{c|c|c|c|c}
          \hline
  \textbf{Digital library} & \textbf{\begin{tabular}[c]{c}Number of studies \\ from the search \\ keywords-based results\end{tabular}} & \textbf{\begin{tabular}[c]{c}Number of studies \\after the selection \\criteria\end{tabular}} & \textbf{\begin{tabular}[c]{c}Number of studies \\ by snowballing\end{tabular}} & \textbf{All} \\\hline
  ACM & 684  & 20 & -- & 20\\\hline
  IEEE& 48 & 32 & -- & 32\\\hline
  Elsevier& 413 & 37 & 2 & 39\\\hline
  Springer & 400 & 42 & -- & 42 \\\hline
  Wiley & 146 & 2 & -- & 2  \\\hline
  PubMed  & 206 & 87 & 5 & 92  \\\hline 
  \textbf{\textit{Total}}&1897	&220	&7	&227\\\hline\hline
  \multicolumn{4}{c|}{\textbf{\textit{After removing 48 duplicate studies}}}&179\\\hline
  \end{tabular}  
  }          
\end{table}

\begin{table}[!t]
  \caption{Data collected for research questions.}
  \centering
  \scriptsize
  \label{tab:data extraction}
   \setlength{\tabcolsep}{4mm}{
          \begin{tabular}{c|m{10cm}}\hline
  \textbf{RQs}& \textbf{Type of extracted data}\\\hline                             
  RQ1 & Number of papers per year and venue type\\\hline
  RQ2 & Existing databases that have been used for RPIP  \\\hline
  RQ3 & Existing techniques for feature encoding\\\hline
  RQ4 & Deep learning algorithms or models\\\hline
  RQ5 & Metrics used for evaluations\\\hline
  RQ6 & Existing DL-based RPIP applications\\\hline
  RQ7 & Existing tools that have been applied for RPIP\\\hline
  RQ8 & Identified open challenges for DL-based RPIP\\\hline   
  \end{tabular}  }          
\end{table}

\section{Answer to RQ1: Publication Situation for DL-based RPIP
\label{SEC:rq1}}

In this section, we address RQ1 by summarizing the primary DL-based RPIP studies according to publication trends and venues.

\subsection{Publication trends}

\begin{figure}[!t]
\centering
\subfigure[Number of publications per year]{
\label{publish (a)}
\includegraphics[width=0.45\textwidth]{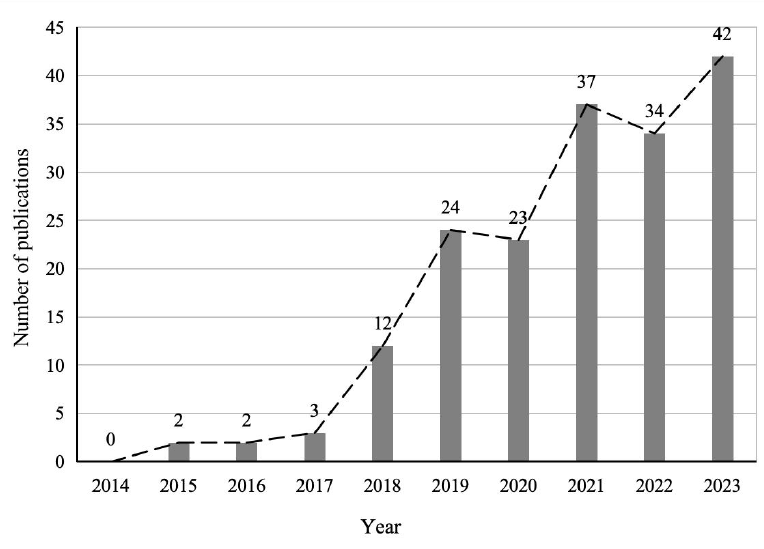}}
\subfigure[Cumulative number of publications per year]{
\label{publish (b)}
\includegraphics[width=0.45\textwidth]{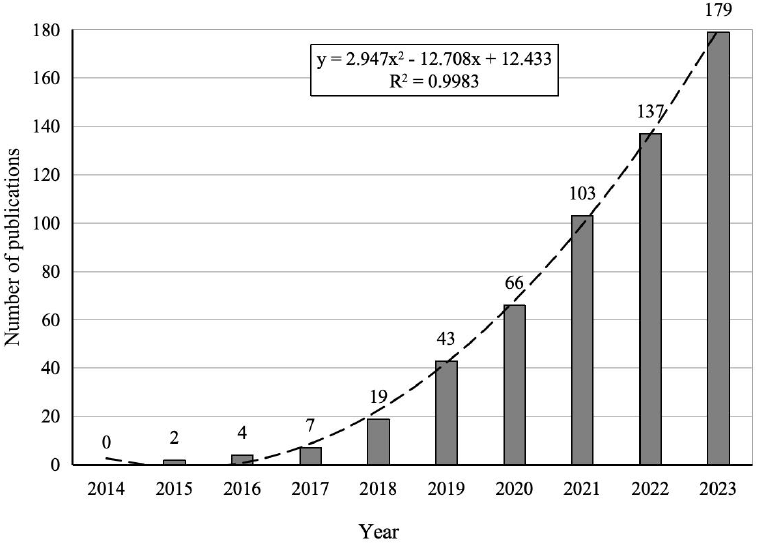}}
\caption{Numbers of DL-based RPIP papers published between January 1st 2014 and December 31st 2023.}
\label{published}
\end{figure}

Figure~\ref{published} presents the DL-based RPIP publication trends from January 1st, 2014, to December 31st, 2023, with Figure~\ref{publish (a)} showing the number of publications per year, and Figure~\ref{publish (b)} showing the cumulative amounts.

It can be observed that no DL-based RPIP paper was published in 2014, with the first articles \cite{alipanahi2015predicting} appearing in 2015.
After that, papers were published every year, increasing substantially in 2018 and peaking in 2023.

An analysis of the cumulative publications (Figure~\ref{publish (b)}) shows that a line function with a high determination coefficient ($R^2 = 0:9983$) can be identified.   
This indicates that the topic of DL-based RPIP has been experiencing strong linear growth, attracting continued interest and showing healthy development.
The 150 (84\%) of the 179 studies under consideration were published in journals, and 29 (16\%) were published in conference proceedings.

\subsection{Publication knowledge maps}

We used VOSviewer~\cite{van2010software} tool to generate co-occurrence, co-authorship, and co-citation maps for the 179 publications under review.
We found that the research focus on DL-based RPIP has demonstrated consistent annual growth over recent years.

\begin{figure}
    \centering
    \includegraphics[width=\textwidth]{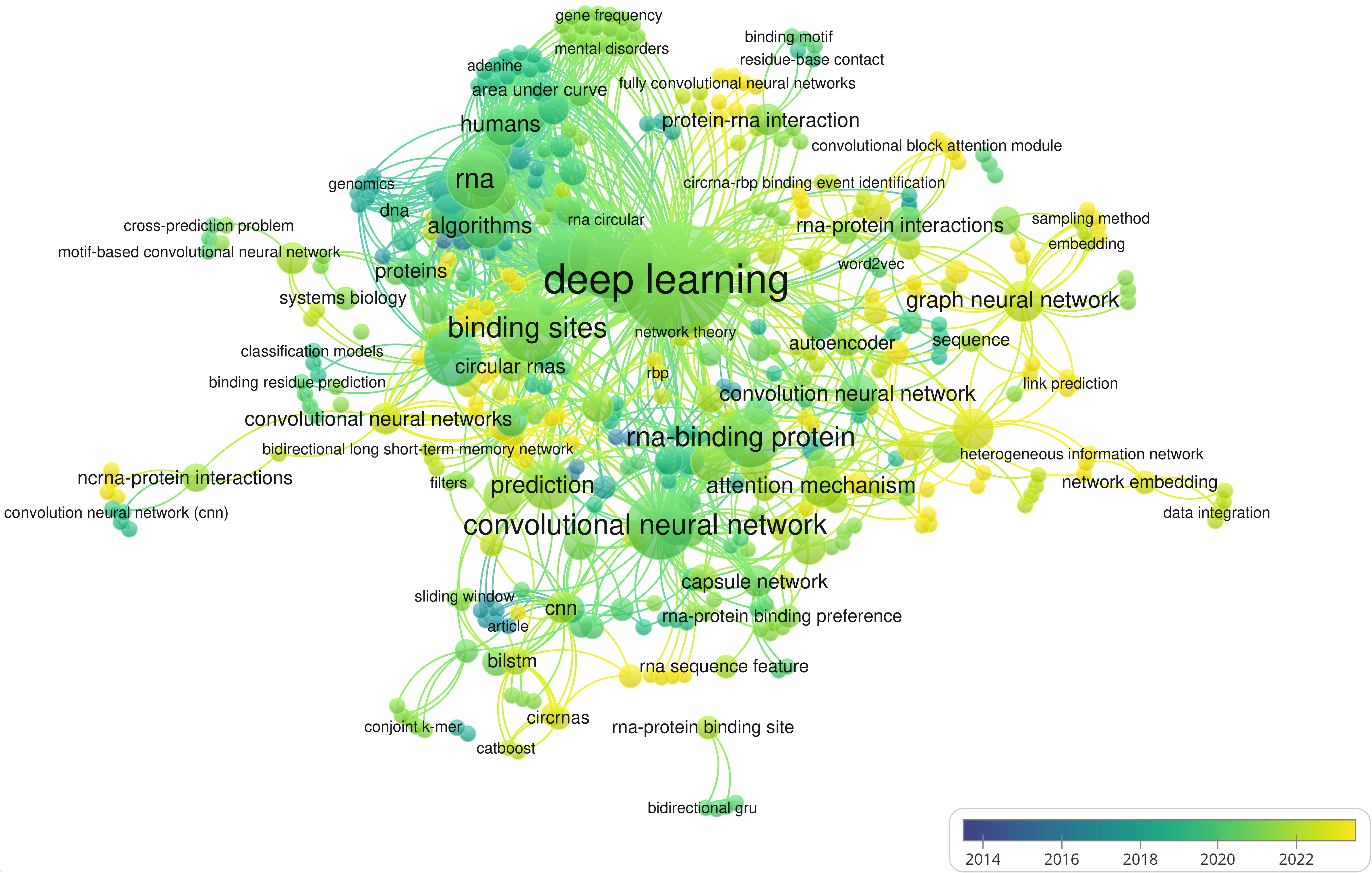}
    \caption{Author-keywords co-occurrence map (nodes represent authors' keywords, and links indicate co-occurrence strength between keywords).}
    \label{FIG: keywords}
\end{figure}

Figure~\ref{FIG: keywords} shows the author-keywords co-occurrence map.
It centers around ``deep learning'', showing the associations among relevant keywords. 
It covers research objects (such as RNA-protein interactions, etc.), technical methods (such as CNN, GNN, etc.), and research focus (such as prediction, etc.), reflecting the associations between DL and various RPIP models, tasks, etc.

\begin{figure}[!t]
    \centering
    \includegraphics[width=\textwidth]{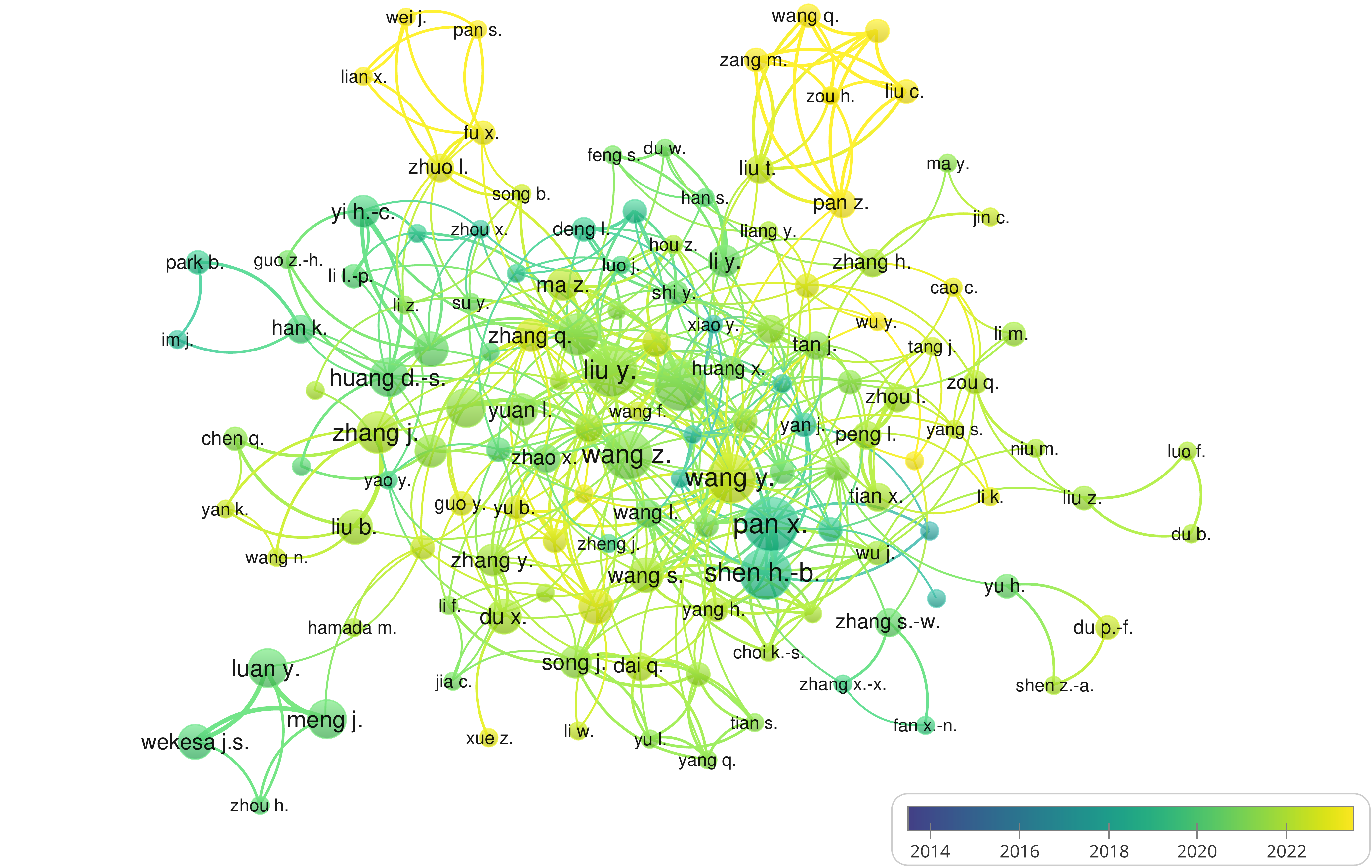}
    \caption{Co-authorship map (nodes represent authors, links indicate collaboration strength, and node size represents publication count).}
    \label{FIG: Authors_co-authorship}
\end{figure}

Figure~\ref{FIG: Authors_co-authorship} shows the co-authorship map.
In this map, nodes represent authors, links denote collaboration strength, and node size indicates publication volume. 
It can be observed that X. Pan, Z. Wang, and Y. Wang each connect multiple scholars, highlighting their dominant research positions.
The densely populated node areas in the figure indicate that collaborators share similar backgrounds, while the extended sparse connections reflect interdisciplinary participation among authors.
The densely connected core primarily reflects theoretical method development (e.g., algorithm design), while long-range connections likely indicate involvement from engineering scholars (applied interdisciplinary collaboration).

\begin{figure}[!t]
    \centering
    \includegraphics[width=\textwidth]{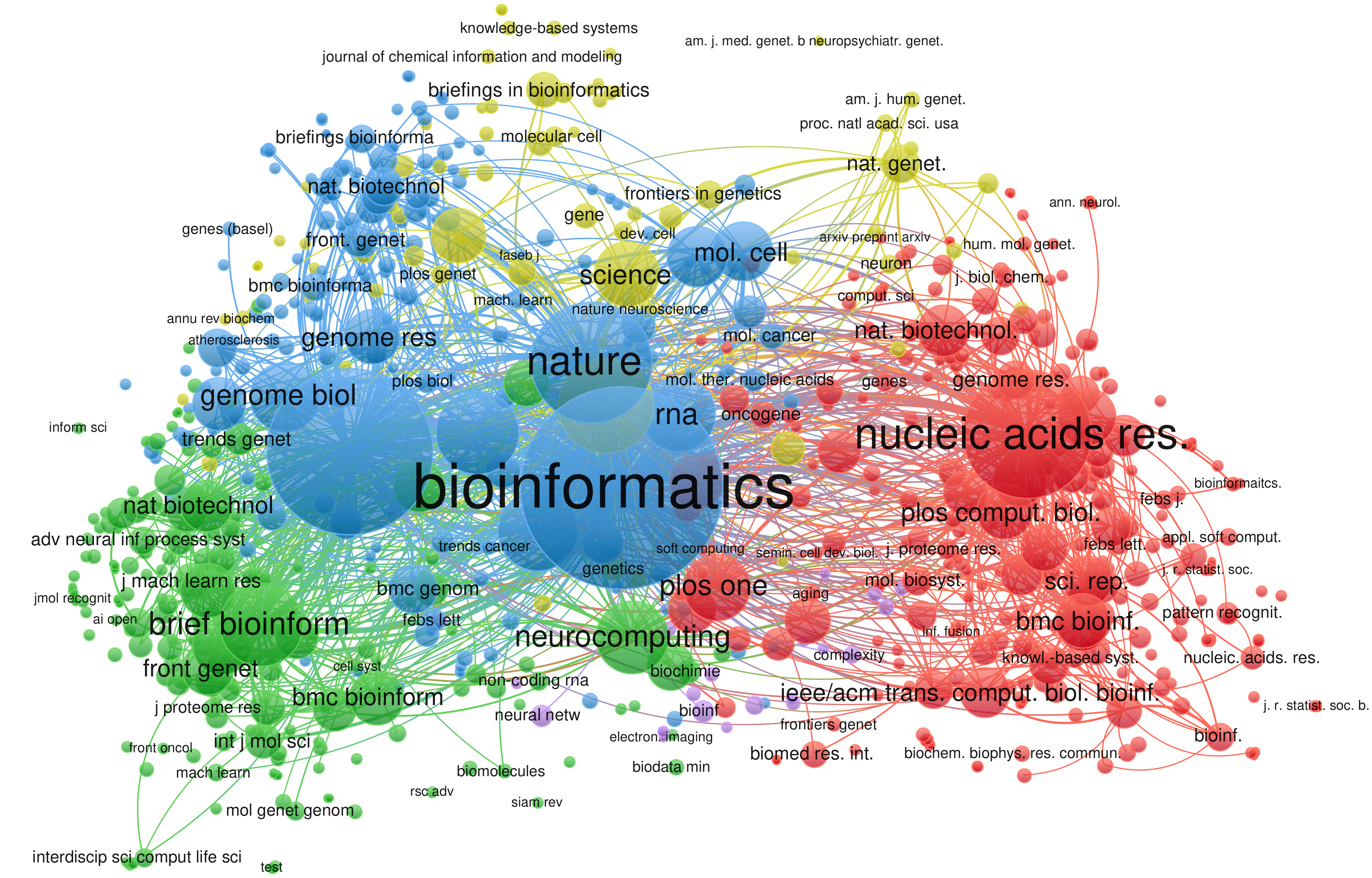}
    \caption{Cited sources co-citation map (nodes represent academic journals or conference proceedings, links show journal pairs co-cited in research papers, and cluster colors denote their broad subject categories).}
    \label{FIG: Cited_sources}
\end{figure}

Figure~\ref{FIG: Cited_sources} shows the cited sources co-citation map.
This map displays nodes as individual publications, links as co-citation relationships, and cluster colors designating research domains (RNA, binding proteins, deep learning, etc.). 
Famous journals like Bioinformatics, Nature, and Science stand out as widely co-cited across disciplines. 
High-impact specialized journals
---
like Briefings in Bioinformatics, Molecular Cell, and Nature Genetics
--- 
bridge methodological and discovery-focused research. 
The colors in the map represent the following:
\textit{Blue:} 
Computational methods (AI algorithms, bioinformatics tools, etc.).
\textit{Red:} 
Cross-technology applications (e.g., computational biomedicine).
\textit{Yellow:}
Multi-Omics analysis.
\textit{Green:} 
Fundamental mechanisms and experimental validation.
Overall, this map validates the deep integration of AI with Biology.

\begin{figure}
    \centering
    \includegraphics[width=\textwidth]{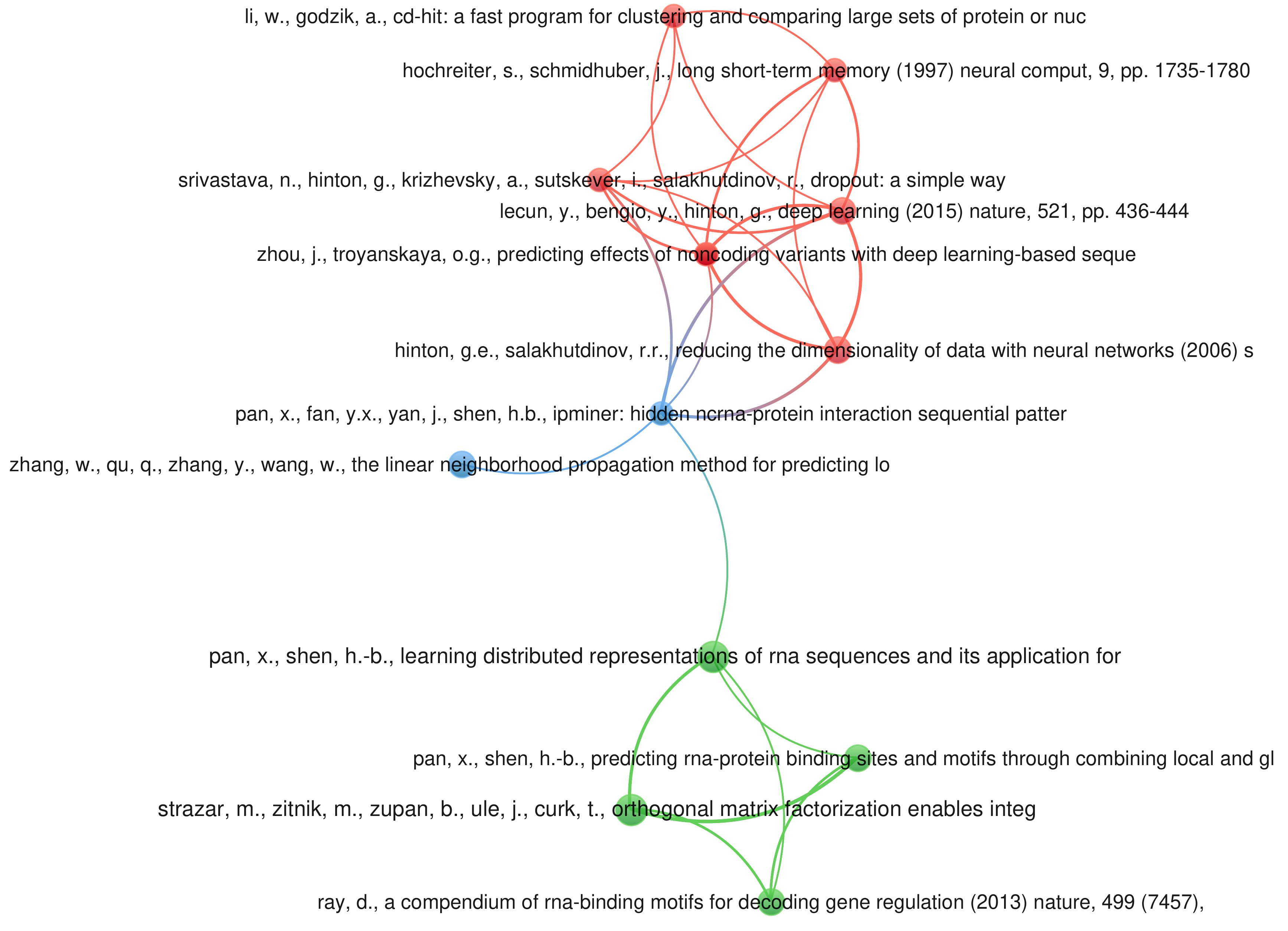}
    \caption{Cited references co-citation map (nodes represent individual publications, links indicate pairs of publications cited together in other sources, and cluster colors reflect research focus similarity).}
    \label{FIG: Cited_references}
\end{figure}

Figure~\ref{FIG: Cited_references} shows the cited references co-citation map.
The citation relationships can provide insights into the important basic literature in the field, the evolution of the research, and the associations among different research achievements.
Figure~\ref{FIG: Cited_references} shows that RPIP and DL are core research areas.
By setting relevant parameters, we display the top 12 most co-cited publications in Figure~\ref{FIG: Cited_references}. 
Three primary research domains can be identified from the article titles:
(1) RNA-protein interaction prediction (represented by Pan et al.);
(2) fundamental DL methodologies (represented by Hinton and LeCun et al.); and
(3) computational biology tools (represented by Li and Godzik's ``CD-HIT'').
Due to display field length constraints in the VOSviewer tool, the textual information in the figure appears truncated:
Complete details of these publications are provided in Table~\ref{tab:references_co-citation}.

\begin{table}
  \caption{Highly co-cited references.}
  \centering
  \scriptsize
  \label{tab:references_co-citation}
  \setlength{\tabcolsep}{1mm}{
          \begin{tabular}{m{2cm}|m{6cm}|m{0.6cm}|m{2.8cm}}
          \hline
  \textbf{Author}  & \textbf{Title} & \textbf{Year} & \textbf{Publication source} \\\hline
  Li et al.~\cite{li2006cd} & Cd-hit: a fast program for clustering and comparing large sets of protein or nucleotide sequences & 2006 & Bioinformatics
  \\\hline
 Hochreiter et al.~ \cite{hochreiter1997long} & Long short-term memory & 1997 & Neural Computation 
  \\\hline
 Srivastava et al.~\cite{srivastava2014dropout} & Dropout: a simple way to prevent neural networks from overfitting & 2014 & The Journal of Machine Learning Research
  \\\hline
 Lecun et al.~\cite{lecun2015deep} & Deep learning & 2015 &  Nature
  \\\hline
 Zhou et al.~\cite{zhou2015predicting} & Predicting effects of noncoding variants with deep learning--based sequence model & 2015 & Nature Methods
  \\\hline
 Hinton et al.~\cite{hinton2006reducing} & Reducing the dimensionality of data with neural networks & 2006 & Science
  \\\hline
 Pan et al.~\cite{pan2016ipminer} & IPMiner: hidden ncRNA-protein interaction sequential pattern mining with stacked autoencoder for accurate computational prediction & 2016 & BMC Genomics
  \\\hline
 Zhang et al.~\cite{zhang2018linear} & The linear neighborhood propagation method for predicting long non-coding RNA--protein interactions & 2018 & Neurocomputing
  \\\hline
 Pan et al.~\cite{pan2018learning} & Learning distributed representations of RNA sequences and its application for predicting RNA-protein binding sites with a convolutional neural network & 2018 & Neurocomputing
  \\\hline
  Pan et al.~\cite{pan2018predicting} & Predicting RNA-protein binding sites and motifs through combining local and global deep convolutional neural networks & 2018 & Bioinformatics
  \\\hline
  Stravzar et al.~\cite{stravzar2016orthogonal} & Orthogonal matrix factorization enables integrative analysis of multiple RNA binding proteins & 2016 & Bioinformatics
  \\\hline
  Ray et al.~\cite{ray2013compendium} & A compendium of RNA-binding motifs for decoding gene regulation & 2013 & Nature
  \\\hline
  \end{tabular}  
  }          
\end{table}

\begin{figure}
    \centering
    \includegraphics[width=\textwidth]{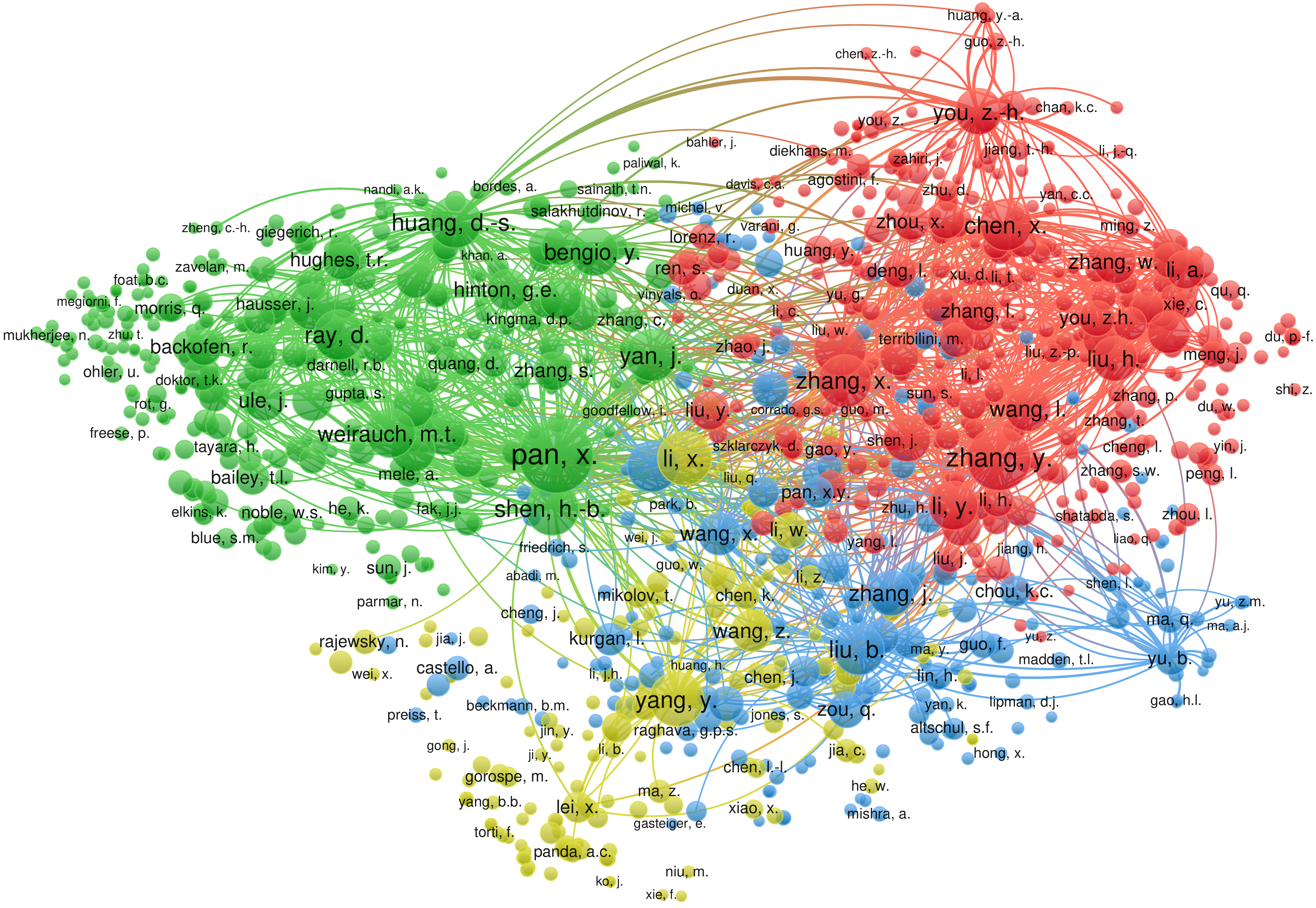}
    \caption{Cited authors' co-citation map (nodes represent authors, links show co-cited author pairs, and cluster colors identify the primary research domains).}
    \label{FIG: Cited_authors}
\end{figure}

Figure~\ref{FIG: Cited_authors} shows the cited authors' co-citation map.
This map shows research-topic affinity among author groups:
Shared cluster colors indicate thematic alignment. 
Links represent co-citation relationships between author pairs: 
Denser and thicker connections signify stronger intellectual influence and scholarly synergy due to frequent joint citations, indicating high-impact knowledge dissemination within the field.

\begin{tcolorbox}[colback=white,sharp corners,boxrule=1pt]
  \textit{\textbf{Summary of answers to RQ1:}}
  \begin{itemize}
    \item[(1)] 
    DL-based RPIP has attracted sustained interest, with the topic showing healthy development. 
    
    \item[(2)] 
    The development of the field can be divided into three stages: 
    The embryonic stage (2014--2017), with an average annual number of publications $<$ 4 papers;
    the acceleration stage (2018--2022), with key technological breakthroughs (such as the application of the Transformer model); and the growth stage (2023), with the annual number of publications continuing to grow.  
    
    \item[(3)] 
    The research community and themes are highly concentrated. 
    Keyword clustering identified the core keywords as:
    ``deep learning'', ``protein-rna interaction'', and ``rna-binding protein''.
    The author co-authorship map and co-citation map show that core scholars (such as Pan and Wang) connect multiple research communities. 
    The sources and references co-citation maps confirm that DL and RPIP analysis constitute core research areas.
    
    \item[(4)] 
    Journals are the main publication venue.
\end{itemize}
\end{tcolorbox}

\section{Answer to RQ2: Datasets that Support RPIP
\label{SEC:rq2}}

The dataset serves as the foundation for training DL models.
In this section, we address RQ2 by summarizing the main datasets in the surveyed literature. 
We first classify the different RPI databases, and then explain how negative samples can be constructed.

\subsection{Related databases}
\label{SEC:RelatedDatabases}

There are many RNA-protein databases. 
According to the 179 surveyed papers, the most commonly used databases were as follows:

(1) Some major public databases storing nucleotide information: 
\textit{Integrated Knowledge Database of Non-coding RNAs} (NONCODE)~\cite{liu2005noncode} and \textit{Database for Circular RNAs} (CircBase) ~\cite{glavzar2014circbase}. 
Gene Ontology~\cite{ashburner2000gene} provides functional annotations for RNA-related genes.
These RNA-related databases contain sequences of ncRNAs, LncRNAs, and circRNAs.
\textit{Nucleic Acid Database} (NDB)~\cite{berman1992nucleic} and RNAshape~\cite{janssen2015rna} contain the structures of RNAs. 
Although \textit{Long Noncoding RNAs Database} (LncRNAdb)~\cite{amaral2011LncRNAdb} is no longer hosted at its original URL\footnote{\url{http://www.LncRNAdb.org/}.}, 
its data is now available in \textit{RNA Central Database} (RNACentral)~\cite{rnacentral2017rnacentral}, currently the most comprehensive database.

(2) Protein databases include \textit{Protein Data Bank} (PDB)~\cite{burley2023rcsb}, \textit{Universal Protein Resource} (Uniprot)~\cite{uniprot2023uniprot}, and \textit{Disordered Proteins Database} (DisProt)~\cite{aspromonte2024disprot}. 
Protein sequences can be obtained from Uniprot and SUPERFAMILY~\cite{gough2001assignment,pandurangan2019superfamily}.
Protein structures are primarily archived in PDB~\cite{burley2022rcsb}; UniProt~\cite{magrane2011uniprot} provides cross-references to PDB entries, while DisProt~\cite{vucetic2005disprot} catalogs disordered regions.
STRING~\cite{szklarczyk2015string} contains protein-protein interaction information.

(3) RPI predictive modeling requires databases that have RPI pairs (labels that can be queried for classification).
RPI databases include the \textit{Non-coding Protein-coding gene Interactions} (NPInter)~\cite{teng2020npinter}, \textit{CircRNA Interactome Database} (CircInteractome)~\cite{dudekula2016circinteractome} and \textit{Protein-RNA Interface Database} (PRIDB)~\cite{lewis2010pridb}.

Many of these databases are not independent, with some being derived from others. 
For example, DisProt also integrates resources from UniProt.
\textit{Circular RNA Database} (CircRNADb)~\cite{chen2016circrnadb} can be linked to CircBase.
STRING relies on other databases, such as PDB and UniProt.
The RPI binding sites in \textit{Encyclopedia of RNA Interactomes} (ENCORI)~\cite{li2014starbase} also include records from CircBase. 
\textit{CLIP-sequencing Database} (CLIPdb)~\cite{yang2015clipdb} is integrated into a submodule of \textit{POST TrAnscriptional Regulation} (POSTAR)~\cite{zhao2022postar3}.
Data in \textit{Database of RNA Interactions} (DoRiNA)~\cite{blin2015dorina} are derived from CLIP-seq. 
Nucleotide sequences were used as inputs to identify cancer-specific circRNA-RBP binding sites.
Although STRING is a protein-protein interaction database, prior knowledge of protein interactions can be embedded into the RPI models~\cite{dai2021pike,jin2023hydra}.
Table~\ref{tab:databases} lists commonly used databases related to RNA, proteins, and their interactions.

\begin{table}
  \caption{Commonly used databases.}
  \centering
  \scriptsize
  \label{tab:databases}
  \setlength{\tabcolsep}{1.5mm}{
          \begin{tabular}{m{1.8cm}|m{4.3cm}|m{6.5cm}}
          \hline
  \textbf{Database}  & \textbf{URL} & \textbf{Brief description}  \\\hline
  CLIPdb~\cite{yang2015clipdb} & \url{http://111.198.139.65/RBP.html} 
  & The CLIPdb provides various annotations for the RBPs, including RNA recognition domains, Gene Ontology, sequence motifs, and structural preferences  
  \\\hline
  PDB~\cite{berman2000protein} &  \url{https://www.rcsb.org/} & Protein Data Bank: Records experimentally determined structures of proteins, nucleic acids, and complexes   
  \\\hline
  NPInter~\cite{yuan2014npinter} & \url{http://bigdata.ibp.ac.cn/npinter5/}  
  & Documents functional interactions between noncoding RNAs and biomolecules 
  \\\hline
  Uniprot~\cite{magrane2011uniprot} &   \url{https://www.uniprot.org/}  
  & UniProt is the world’s leading high-quality, comprehensive, and freely accessible resource of protein sequence and functional information
  \\\hline
  CircInteractome ~\cite{dudekula2016circinteractome} & \url{https://circinteractome.nia.nih.gov/} & Circular RNA Interactome: Regulatory information between circular RNA, RBP, and \textit{MicroRNA} (miRNA)
  \\\hline
  PRIDB~\cite{lewis2010pridb} & \url {http://bindr.gdcb.iastate.edu/PRIDB} 
  & The protein-RNA interface database
  \\\hline
  NONCODE ~\cite{liu2005noncode} & \url{http://www.noncode.org/index.php} 
  & An integrated knowledge base for the study of non-coding RNAs (excluding transfer RNA and ribosomal RNA)
  \\\hline
  NDB~\cite{lawson2024nucleic} & \url{https://nakb.org/} 
  & Repository for experimentally determined nucleic acid and nucleic acid-containing complex structures  
  \\\hline
  ENCODE~\cite{sloan2016encode} & \url{https://www.encodeproject.org/} 
  & Comprehensive resource of functional genomic annotations (e.g., RBP binding) in human and model organisms
  \\\hline
  PlncRNADB ~\cite{bai2019PlncRNADB} & \url{ https://bis.zju.edu.cn/PlncRNADB/index.php} 
  & Plant long noncoding RNA Database 
  \\\hline
  \end{tabular}  
  }          
\end{table}

\subsection{Dataset attributes}

\begin{table}
  \caption{Commonly used datasets.}
  \scriptsize
  \centering
  \label{tab:datasets}
  \setlength{\tabcolsep}{0.6mm}{
  \begin{tabular}{m{1.2cm}|m{1.5cm}|m{1cm}|m{1.2cm}|m{2cm}|m{5.5cm}}\hline
  \textbf{Datasets}&\textbf{Data provenance}& \textbf{Contains negative samples}&\textbf{RNA types}&\textbf{Species coverage}&\textbf{Number of samples}
  \\\hline                       
  RBP24 ~\cite{maticzka2014graphprot} & ENCODE, GEO  &  Yes & mRNA, LncRNA, miRNA, snoRNA, tRNA & Homo sapiens  & The RBP24 dataset contains 24 types of RBPs
 \\\hline
  RBP31 ~\cite{pan2018prediction} & POSTAR  & Yes  & mRNA, LncRNA  & Homo sapiens  &  The dataset contains binding data for 31 types of RBPs
 \\\hline
  RBP37 ~\cite{ray2017rnacompete}&  RNAcompete & No clearly defined negative samples  &  Synthetic random RNA & -  &  Provide a binding affinity matrix for each RBP. Data not derived from biological samples, but from high-throughput screening
 \\\hline
  RPI369 ~\cite{muppirala2011predicting}  & PDB, PRIDB  & No  
  &  mRNA, ncRNA &  Multi-species & It consists of 332 RNA sequences, 338 protein sequences, and 369 positive pairs  
 \\\hline
  RPI488 ~\cite{pan2016ipminer} & PDB, PRIDB   &  Yes & ncRNA &  Homo sapiens / Mouse &  It contains 245 negative RNA-protein pairs and 243 interacting RNA-protein pairs 
 \\\hline
  RPI1807 ~\cite{suresh2015rpi} &  NDB, PRIDB &  Yes & LncRNA &  Homo sapiens / Mouse &  It contains 1436 negative LncRNA-protein pairs and 1807 interacting LncRNA-protein pairs  
 \\\hline
  RPI2241 ~\cite{muppirala2011predicting}  & PDB, PRIDB  & No  
  & mRNA, LncRNA  &   Multi-species & It contains 2,241 positive interaction pairs generated from 943 protein RNA complexes extracted  
 \\\hline
  RPI7317 ~\cite{hao2016npinter} & NPInter3  &  No 
  & LncRNA, mRNA &  Homo sapiens & It contains 7317 RNA-protein interaction pair samples  
 \\\hline
  NPInter 10412 ~\cite{yuan2014npinter} &  NPInter2 & No  
  & ncRNA & Multi-species  &  It contains 10,412 RNA-protein interaction pair samples   
 \\\hline
  ATH948 ~\cite{bai2019PlncRNADB}  & PlncRNADB  & No  &  LncRNA & Arabidopsis thaliana  & It contains 948 interactions between 390 LncRNAs and 163 proteins  
 \\\hline
  ZEA22133 ~\cite{bai2019PlncRNADB}  &  PlncRNADB & No  
  & LncRNA &  Zea mays &  It contains 22133 interactions between 1107 LncRNAs and 190 proteins      
 \\\hline
  \end{tabular}  
  }
\end{table}

Using screening, splicing, and redundancy-removal, various datasets have been constructed from the databases in Section~\ref{SEC:RelatedDatabases}.
Table~\ref{tab:datasets} lists the datasets frequently used in the surveyed literature.

If the number of positive RPI pairs is $x$, then the datasets are named RPI$x$ or RBP$x$.
In Table~\ref{tab:datasets}, RPI488, RPI1807, RPI2241, and RPI369 are structure-based. 
According to the minimum atomic distance, if the distance between a protein atom and an RNA atom is less than the specified threshold, the protein and RNA are considered to be an interaction pair. 
For example, RPI1807 sets 3.4~\AA ~as the threshold between positive and negative interactions. 
NPInter2.0 is a non-structure-based dataset that primarily contains experimentally validated data.
RBP24 and RBP31 were obtained through CLIP-seq processing with DoRiNA.
Some studies filter their own datasets through databases or generate their private datasets through biological experiments~\cite{ben2018deep,zhao2021multi}.
Additionally, negative samples may be defined using data distribution analysis and prior knowledge. 
The interaction scores of high-confidence interaction subsets serve as the threshold~\cite{navamajiti2019mcbel,arora2022novo}.

RPIs are usually represented as RPI pairs, whereas RBPs often provide information on binding regions in addition to representing RNA binding protein pairs.
However, as highlighted in Table~\ref{tab:datasets}, some datasets do not contain negative samples.

\subsection{Dataset-bias analysis}

Mainstream RPI datasets may have several systemic biases.
\textbf{Species bias:} 
Human-derived data dominates (e.g., RBP24: 87\%; NPInter10412: $>$95\%), while coverage of other species, especially plants (e.g., ATH948, ZEA22133), remains inadequate.
\textbf{RNA-type imbalance:} 
NPInter10412 disproportionately emphasizes noncoding RNAs.
\textbf{Data quality issues:} 
Smaller datasets like RPI488 can have significant noise. 
RPI2241 lacks standardized deduplication with inconsistent positive/negative sample ratios.
\textbf{Technical bias:} 
Hybridization of experimental datasets (e.g., CLIP-seq-dominated RBP31) and purely computational datasets (e.g., RPI2241) may introduce methodological artifacts.

Although current RPI datasets may have selection biases, appropriate datasets can still be selected based on specific tasks or model requirements.
RPI datasets represent interacting pairs, and RBP datasets provide binding regions of RNA-binding proteins.
Although there is no universal standardized benchmark, domain-specific benchmarks have been widely adopted. 
Researchers should clarify their objectives when selecting task-matched datasets:
RPI$x$ datasets should be used for interaction-pair prediction; while 
RBP$x$ datasets would be suitable for binding site prediction. 
Published studies typically undergo testing on more than three public datasets and follow multi-metric evaluation, enabling fair comparisons.
Research papers or reproducibility kits should be provided alongside the code to further enable direct performance comparisons of DL algorithms.

\subsection{Construction of negative samples}

Most of the data provided in the databases consists mainly of positive samples (interactions exist):
This leads to a lack of negative samples (no interactions), resulting in imbalanced datasets. 

The negative samples in RPI1807 (in Table~\ref{tab:datasets}) were determined using a threshold greater than 3.4~\AA.
RBP24 generated negative sites by shuffling the coordinates of at least one binding site in all genes using \textit{Browser Extensible Data Tools} (BedTools)~\cite{quinlan2010bedtools}.
The RBP31 negative samples were collected from positions that had not been validated as binding sites.
The RPI488 negative dataset was based on the minimum atomic distance. 

Various randomness-related methods were used on datasets lacking negative samples.
The negative samples in RPI369 and RPI2241 were all randomly generated.
The non-interacting pairs in \textit{Arabidopsis thaliana} and \textit{Zea mays} datasets (ATH948 and ZEA22133) were generated by randomly pairing proteins with RNA, and by removing some positive pairs.

Construction of negative datasets can be categorized into the following approaches:
\begin{itemize}
    \item 
    \textbf{Flanking region truncation:}
    The negative datasets are sampled from upstream and downstream of the RNA protein interaction region ~\cite{shen2020deep,wang2021prediction}.
    
    \item 
    \textbf{Remaining choices:} 
    After removal of the interaction sites, negative samples are taken from the remaining sequences ~\cite{li2021capsule,peng2022enanndeep}.
    
    \item 
    \textbf{Random sampling:} 
    This involves randomly selecting negative samples from unverified RNA protein interaction pairs~\cite{peng2021finding,zhao2022predicting}.
    
    \item 
    \textbf{Indirect search:} 
    Searching databases or other literature, and filtering through annotation libraries~\cite{zhao2020econvrbp,qiu2020prona2020}.
    
    \item 
    \textbf{Interaction scoring threshold:}
    Interaction scores of RNA-protein pairs are calculated based on known RPIs. 
    Scores below the threshold are considered to be negative samples~\cite {huang2022lpi,zhang2020lpi}.
    
    \item 
    \textbf{Space structures:} 
    Similar to the RPI488 construction approach, negative samples are identified when the distance between the RNAs and proteins exceeds the threshold ~\cite{zhao2022dfpin,li2022pst}.

    \item 
    \textbf{Sequence variant reconstruction:}
    Similar to the RBP24 construction approach, negative samples are generated by modifying or shuffling positive sequences~\cite{alipanahi2015predicting,wang2022web}.
\end{itemize}

Most DL models perform well on balanced datasets; however, how to sample or enhance negative data is a question worth exploring more fully (as we discuss in Section~\ref{SEC:rq8}).

\begin{tcolorbox}[colback=white,sharp corners,boxrule=1pt,breakable]
    \textit{\textbf{Summary of answers to RQ2:}}
    \begin{itemize}
        \item[(1)] 
        RPI databases have a three-tier classification hierarchy: 
        RNA-specific databases, protein-specific databases, and dedicated RPI databases. 
        Prevalent cross-database dependencies necessitate validation across multiple sources to ensure data consistency.
        
        \item[(2)] 
        Systemic biases persist in mainstream datasets. 
        Optimal dataset selection should align with task-specific requirements, such as using RBP24 for sequence motifs and RPI1807 for interaction-pair prediction.  
        
        \item[(3)] 
        Negative sample construction methodologies significantly impact model robustness.
        
       \item[(4)] 
       Standardized evaluation benchmarks are urgently needed. 
       It is necessary to establish cross-database benchmarks, and to develop dual-track evaluation frameworks for experimental and computational datasets.
    \end{itemize}
\end{tcolorbox}

\section{Answer to RQ3: Feature Extraction and Encoding of RNA and Protein Information
\label{SEC:rq3}}

Before processing can take place, the raw RNA and protein data need to be converted into a structured input format that DL models can process.
A major challenge for RPIP is the need to design an effective representation for RNA/protein sequences and structures, enabling their transformation into fixed-dimensional feature vectors. 
Feature-encoding results can be obtained from several tools, such as \textit{Amino Acid Index}~\cite{kawashima2007aaindex}, \textit{Python-based Effective Feature Generation Tool} (Pyfeat)~\cite{muhammod2019pyfeat}, \textit{Biomolecule Tri-class Angle Platform for Integrated Exploration} (BioTriangle )~\cite{dong2016biotriangle}, and \textit{Pseudo components in One-stop shop} (Pse-in-One)~\cite{liu2015pse}.
This section addresses RQ3 by explaining the sequence and structural information of RNAs/proteins, and discussing the most commonly used feature and encoding methods.

\subsection{Sequence features}

RNA and protein features can be divided into sequences and structures. 
The sequence features are classified as primary sequences, constituent components, physicochemical properties, and evolutionary information.

\textit{$\bullet$ Primary sequence features and encoding}

The primary RNA sequences refer to nucleotide sequences, and the primary protein sequences refer to amino acid sequences. 
The most commonly used method for encoding primary sequences is one-hot encoding~\cite{pan2018learning}, which treats the sequence as a string of characters.
Although this method is simple, it can result in high-dimensionality and sparsity problems~\cite{du2022deep,cui2022deepmc}.
Using high-order one-hot encoding may relieve sparseness, but it does not have a significant effect~\cite{zhang2020lpi,zhang2019prediction}.
The $k$-mer encoding~\cite{guo2022pseudo} is another popular method:
The $k$-mer frequency refers to dividing a sequence into subsequences of length $k$, and then counting the number of these subsequences~\cite{shen2021npi,zhuo2022predicting}.
Gapped $k$-mer (allowing interspersed bases), reverse $k$-mer (including reverse complementary sequences), and hybrid nucleotide frequencies (incorporating mixed nucleotide and structural features) are extensions of the $k$-mer frequency approach~\cite{tahir2021kdeepbind,huang2022ncrna,zhang2021predicting}.

A word-embedding approach can also be used to encode sequences.
The $k$-mer frequency vectors are similar to bags of words in natural language processing.
Therefore, subsequences divided by $k$-mers can also be embedded into word vectors using \textit{Word to Vector} (word2vec) or \textit{Document to Vector}‌‌ (doc2vec).
This technique encodes biological sequences, considers each $k$-mer as a word or phrase in a sentence, and uses the word2vec algorithm (continuous bag of words or skip-gram) to learn distribution representations to train word-embedding models~\cite{deng2019deep,pan2018learning}.

RNA sequences have only four types of nucleotides.
One-hot and $k$-mer encoding methods are commonly used. 
\textit{Biological Vectors}~\cite{jin2021predicting}, \textit{RNA to Vectors} (RNA2vec)~\cite{ju2019circslnn}, and \textit{circRNA to Vectors} (circRNA2Vec)~\cite{du2022jlcrb} are often used for RNA word embedding.
However, proteins have 20 amino acids, so using one-hot encoding directly could be very complex. 
Therefore, for protein sequences, $k$-mer is used more frequently.
The \textit{Protein to Vectors} (Pro2vec)~\cite{yi2020learning} word-embedding model is also an effective encoding method.

\textit{$\bullet$ Component features and encoding}

The component features represent higher-level granularity information.
RNA dinucleotides and trinucleotides are the basic components of RNA molecules, while protein dipeptides and tripeptides are short peptide-chain segments in long protein chains.
RNAs and protein molecules are usually encoded based on the frequency of their composition, such as nucleotide density~\cite{yang2022hcrnet}, trinucleotide composition~\cite{liu2023crbsp}, and composition-transition-distribution~\cite{du2018deepmvf,zheng2018deep}.
However, if only the frequency is used, then some positional information may be missed:
Therefore, some methods increase positional information
---
such as order preservice transformation~\cite{wang2018combining,wang2019prediction}, pseudo amino acid composition~\cite{fan2019lpi,wang2023prediction}, and position-specific trinucleotide propensity~\cite{liu2023crbsp}
---
to better represent and encode the compositional characteristics of RNAs and protein molecules.

\textit{$\bullet$ Physicochemical features and encoding}

The physical and chemical properties can affect the structure and function of molecules. 
Tools for encoding the physicochemical feature of RNAs and proteins include Pyfeat (which extracts features from thirteen different techniques)~\cite{muhammod2019pyfeat}
and BioTriangle (which calculates nine types of features)~\cite{zhou2021lpi,zhou2021lpib}.
The encoded features include dozens of physical and chemical characteristics, such as static electricity, polarity, hydrophobicity, and hydrophilicity.

\textit{$\bullet$ Evolutionary features and encoding}

The evolutionary features of RNAs and proteins involve complex changes in structure, function, and other aspects, affecting the molecules' interactions.
Encoding the evolutionary features involves the extraction of useful evolutionary information to represent the RNAs and proteins.

The \textit{Position Specific Scoring Matrix} (PSSM) and \textit{Position Specific Frequency Matrix} (PSFM), and their variants, are the most commonly used encoding methods for protein evolutionary information~\cite{shibu2023prediction,zhang2020idrbp_mmc}. 
Although \textit{Position-Specific Iterated Basic Local Alignment Search Tool} (PSI-BLAST) generates both PSSMs and PSFMs, they extract evolutionary information in different ways. 
Multiple sequence alignment can be used to represent proteins' evolutionary information~\cite{roche2024equipnas}.

The RNA evolutionary information and encoding methods are relatively limited, but include conservation scores~\cite{tayara2020improved,vaculik2023transfer}, motif information~\cite{li2021capsule}, and \textit{Hierarchical Hidden Markov Model-based Blind  Iterative Threading Searcher} (HHblits), which uses hidden Markov models to search for distant homologs~\cite{jiang2023structure}.

\subsection{Structure features}
Due to the time and resource requirements of biological experiments, the number of known RNA and protein structures is relatively limited.
However, with the increasing applications of AI in biological science, some tools can predict the structures of proteins and RNAs, which can be encoded into vectors as the features of RNAs and proteins.

\textit{$\bullet$ RNA-structure features and encoding}

RNA structures influence protein-binding sites. 
Because sequence conservation is low among distantly related species, but structural conservation remains high, using only sequence features could result in structural constraints being overlooked~\cite{johnsson2014evolutionary}. 
Structure-prediction tools computationally model static folding patterns to identify RBP-binding motifs~\cite{lorenz2011viennarna}. 
Furthermore, integrating structural features with sequence features within a multi-modal framework enhances model performance~\cite{baek2024accurate}.
Structure-prediction tools partially alleviate the scarcity of experimentally determined structural data, providing spatio-structural information for DL models.
RNA-structure prediction tools include RNAfold~\cite{mohsen2009comparison}, which is one of the core programs of the Vienna package~\cite{gruber2008vienna}:
It can be used to predict the secondary structures of single sequences using dynamic programming algorithms~\cite{yang2018lncadeep,liu2021rna}.
RNAshapes \cite{pan2018prediction,sun2021predicting} transforms RNA structures into a new sequence composed of six types of RNA secondary structures.
\textit{Single-sequence-based Prediction Of Tertiary RNA structures} (SPOT-RNA) is an RNA secondary structure prediction method that uses seven single-character identifiers to represent the structural type of each nucleotide~\cite{singh2019rna}. 
\textit{Java-based Alignment of RNA using 3 Dimensional Structure} (JAR3D) finds possible 3D geometries for hairpin and internal loops using an RNA 3D motif atlas~\cite{zhang2016deep}.

\textit{$\bullet$ Protein-structure features and encoding}

AlphaFold is currently the most advanced tool for protein structure prediction~\cite{roche2024equipnas}.
\textit{Define Secondary Structure of Proteins} (DSSP)~\cite{gorelov2024dssp} assigns secondary structures to amino acid residues based on 3D atomic coordinates from PDB files. 
\textit{Self-Optimized Prediction Method with Alignment} (SOPMA) predicts secondary structures from protein sequences using self-optimized alignment methods~\cite{geourjon1995sopma}.
SPIDER calculates the secondary structures of proteins and can also predict the tertiary structure~\cite{liu2021aprbind}. 
\textit{Protein Secondary Structure Predictor} (SSPro) predicts the secondary structures of proteins~\cite{huang2022lpi,wekesa2020deep}.

Although the tools described so far can predict RNA and protein structures, they do not all output numerical values:
RNAshapes transforms RNA structures into a new sequence composed of six types of RNA secondary structures (\textit{Stem}, \textit{Multiloop}, \textit{Hairpin}, \textit{Internal loop}, \textit{Dangling end}, and \textit{Dangling start}). 
SPOT-RNA uses seven single-character identifiers to represent the structural type of each nucleotide in the primary sequences. 
SOPMA divides the secondary structure of protein-sequence residues into three categories (alpha-helix, beta-sheet, and coil). The non-numerical output from tools like these must be converted into vectors: 
Structural characters are serialized and encoded into vectors that can be used as DL model inputs.
Simultaneously, the converted secondary structure features can be captured by learnable vector representations to reveal latent structural semantics. 
Furthermore, they enable feature fusion with sequence features, in a unified feature space.
Conversion of the structure features into character sequences means that sequence-encoding methods can then be used.
The secondary structure features are represented as sequences and encoded using one-hot, $k$-mer frequency, or other methods.

\subsection{Feature-encoding analysis}
Our survey of the literature revealed that although RNA and protein sequences and structures are often used together, few studies have examined which type of features and encoding methods are the best~\cite{wekesa2020multi}. 
During feature selection for RNA and protein representations, contribution balancing in feature fusion can be achieved through dynamic weighting mechanisms or optimization algorithms, as shown by Khatir et al.~\cite{khatir2024efficient, khatir2023new}:
Compared with purely data-driven models, their framework maintains enhanced prediction stability even with limited data, thereby significantly improving model robustness.
Dynamic weighting and optimization methods adaptively adjust multi-source feature contributions while optimizing their combined effect through intelligent searches. 
By assigning weights to the input features, these methods can model complex nonlinear relationships between features and outputs. 
Exploration-exploitation mechanisms dynamically narrow the solution space toward the global optimum. 
Finally, multi-objective optimization balances the feature fusion, and enhances noise robustness.

\begin{figure}
    \centering
    \includegraphics[width=0.6\textwidth]{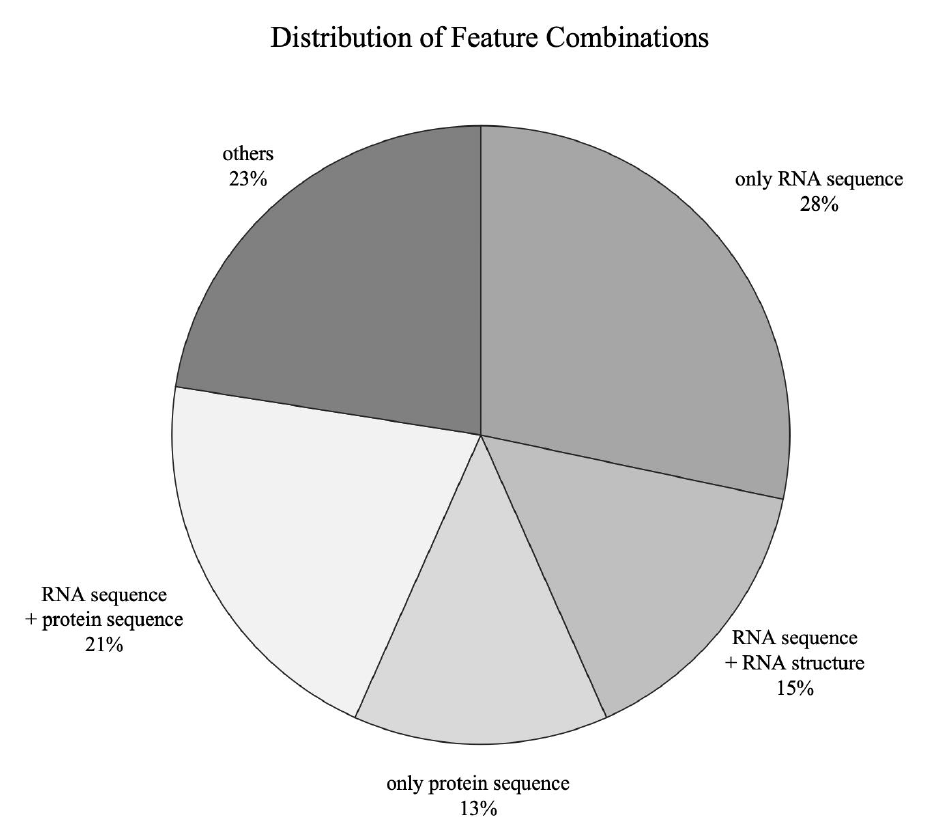}
    \caption{Distribution of feature combinations in the surveyed literature (the ``others'' category represents additional combinations of RNA and protein sequence/structure features).}
    \label{FIG:feature}
\end{figure}

We quantified the utilization of RNA and protein sequence/structure features in the 179 surveyed publications:
Figure~\ref{FIG:feature} presents the resulting feature-combination distribution (only combinations with usage proportions exceeding 10\% are shown). 
In Figure~\ref{FIG:feature}, ``others'' represents other feature combinations of RNAs and proteins (such as using only protein structures, only RNA structures, or only the adjacency matrix of RPI graph).
It can be seen from the figure that sequence features of RNAs and proteins have been widely used.
Notably, only one method was based on text processing~\cite{li2016text}, using a text-feature-based text-mining approach to extract RPI from the biomedical literature.
This method first collects literature data from \textit{PubMed}, \textit{LncRNAdb}, and the \textit{LncRNA Disease Database}, unifying text formats through Python scripts. 
Text preprocessing is then performed:
This involves deduplication, sentence segmentation, named entity recognition, part-of-speech tagging, and shallow syntactic parsing to generate grammatical chunks. 
Next, five categories of text features are extracted:
lexical features capture syntactic roles around entities;
phrase features detect predefined trigger words;
verb features analyze interaction semantics;
syntactic features parse dependency structures; and
auxiliary features determine entity attributes.
Finally, binary classification is performed on the feature vectors to identify authentic RPI descriptions, outputting high-confidence interaction pairs.
This approach faces challenges, however, including domain-specific term ambiguity and noisy data, preventing it from becoming mainstream.

\begin{table}
    \centering
    \caption{Feature encoding comparison.}
    \label{tab:feature encoding}
    \begin{threeparttable}
    \scriptsize
    \setlength{\tabcolsep}{1.5pt} 
    \begin{tabular}{
        >{\flushleft\arraybackslash}m{2cm}|
        >{\raggedright\arraybackslash}m{2.2cm}|
        >{\raggedright\arraybackslash}m{3cm}|
        >{\raggedright\arraybackslash}m{3cm}|
        >{\raggedright\arraybackslash}m{2.5cm}}
        \hline
        \textbf{Features} & \textbf{Method} & \textbf{Advantages} & \textbf{Limitations} & \textbf{Applicable scenarios} \\
        \hline
        \multirow{6}{1.1cm}{Primary sequence} 
        & One-hot encoding  & Simple and interpretable  & High-dimensional sparsity & Rapid prototyping \\
        \cline{2-5}
        &$K$- mer frequency encoding &  Captures local sequence patterns & Insufficient long-range dependencies & Rapid prototyping, binding site prediction  \\
        \cline{2-5}
        &Word embedding & Models semantic associations & Dependent on big data pre-training & Sequence representation learning  \\
        \cline{2-5}
        \hline
        \multirow{4}{1.1cm}{Component features}
        & Nucleotide/Amino acid composition frequency & Simple, efficient, reflects compositional trends & Ignores sequence order and positional information & Coarse-grained candidate screening \\
        \cline{2-5}
        & Pseudoamino acid composition& Integrates physicochemical properties and sequence order information & Complex parameter optimization & Functional domain prediction \\
        \cline{2-5}
        \hline
        \multirow{3}{1.1cm}{Physicochemical features}
        & Pyfeat &  Comprehensive molecular properties & Feature redundancy requires reduction & Multidimensional interaction analysis \\
        \cline{2-5}
        & BioTriangle & Calculate multiple physicochemical indices & Limited predictive power of single features & Multidimensional interaction analysis \\
        \cline{2-5}
        \hline
        \multirow{5}{1.1cm}{Evolutionary features}
        &PSSM/PSFM &  Preserves site conservativeness & Relies on multiple sequence alignment, time-consuming & Identification of key functional domains \\
        \cline{2-5}
        & RNA conservation score & Reflects structural/functional constraints & Sparse data for non-coding RNAs & Functional annotation \\
        \cline{2-5}
        & HHblits & Detects distant homologous sequences & Computationally intensive & Identifying cross-species RBP homologs \\
        \cline{2-5}
        \hline
        \multirow{5}{1.1cm}{Structure features}
        & RNAfold/SPOT-RNA &  Preserves stem-loop topology, overcoming limitations of sequence-only analysis, identifying functional motifs, enabling multi-modal integration & Limited accuracy on tertiary interactions & RNA secondary structure dependency analysis \\
        \cline{2-5}
        & DSSP/SOPMA & Provides protein secondary structural information & Lacks conformational dynamics & Protein secondary structural features  \\
        \cline{2-5}
        & AlphaFold & High-accuracy structural features & Training data bias, black-box representation & Provides 3D structural features of proteins \\
        \cline{2-5}
        \hline
    \end{tabular}
    \end{threeparttable}             
\end{table}

The main objective of feature encoding in DL-based RPIP is to transform biological sequences and structural information into fixed-dimensional numerical vectors.
Table~\ref{tab:feature encoding} presents some common encoding methods. 
Sequence encoding focuses on capturing local patterns and semantic associations, but requires balancing the trade-offs between high-dimensional sparsity and data dependency. 
Compositional feature encoding integrates physicochemical properties with sequence order, making it suitable for functional domain prediction. 
Physicochemical feature tools provide multidimensional molecular attribute descriptions, but require dimensionality reduction to be able to handle redundancy. 
Evolutionary features capture functional constraints through evolutionary conservation and remote homology detection, but they depend on time-consuming alignment procedures and currently face data scarcity in ncRNA datasets. 
Structural encoding uses secondary topology and spatial conformation to localize binding domains, but remains limited for modeling tertiary interactions (e.g., pseudoknots) and dynamic conformational changes.

Compared with traditional feature encoding, biological language models learn deep semantic patterns in biological sequences through self-supervised pretraining, providing cross-dimensional support for RPIP. 
Their pretraining-finetuning paradigm can transform RPIP feature-representation learning. 
Transformer-based biological pretrained models (e.g., \textit{DeoxyriboNucleic-Acid Bidirectional Encoder Representation from Transformers} (DNABERT)~\cite{ji2021dnabert} and \textit{Protein Transformer} (ProtTrans)~\cite{elnaggar2021prottrans} directly generate context-aware embeddings from raw RNA/protein sequences, replacing traditional encoding methods and significantly enhancing cross-task generalization capabilities. 
Finetuned language models, like attention mechanisms incorporating positional encoding, can predict RBP binding propensity or binding sites. 
Self-attention weights can localize key residues/nucleotides (e.g., high-weight regions corresponding to conserved binding sites), offering potential biological mechanistic explanations for ``black-box models''.
However, biological language models cannot fully replace structural feature encoding. 
They also face other challenges, including dependence on large-scale pretraining data, insufficient integration of structural information, and high computational costs. 
Traditional encoding remains indispensable in small-sample or resource-constrained scenarios. 
Future research directions may focus on multimodal language models (jointly pretraining sequence-structure-interaction networks) and generative applications.

\begin{tcolorbox}[colback=white,sharp corners,boxrule=1pt]
    \textit{\textbf{Summary of answers to RQ3:}}
    \begin{itemize}
        \item[(1)] 
        Sequence feature-encoding involves a dimensionality-information content trade-off.
        One-hot encoding is simple and intuitive, but leads to high-dimensional sparsity.
        $K$-mer encoding preserves local conserved motifs, but loses long-range dependencies.
        Word embeddings model semantic relationships, but rely on large-scale pretraining.
        
        \item[(2)] 
        Compositional features require positional information enhancement.
        Mature feature-extraction tools exist, such as Pyfeat and BioTriangle.
        
        \item[(3)] 
        Evolutionary feature-encoding exhibits species asymmetry, with sparse RNA evolutionary features.
        
        \item[(4)] 
        Structural feature-encoding relies on a cascaded structure $\rightarrow $ sequence $\rightarrow$ vector encoding paradigm.
        
        \item[(5)] 
        Key research directions include multimodal feature fusion and cross-disciplinary methodological convergence.
    \end{itemize}
\end{tcolorbox}

\section{Answer to RQ4: Model-Construction Techniques Used by DL-based RPIP
\label{SEC:rq4}}

Encoded or transformed RNA/protein data serve as inputs to the DL models, which can learn the complex input-output mappings.
This section provides an overview of the applications of DL models in RPIP, examining them from three dimensions: basic, novel, and hybrid models.

\begin{figure}
    \centering
    \includegraphics[width=\textwidth]{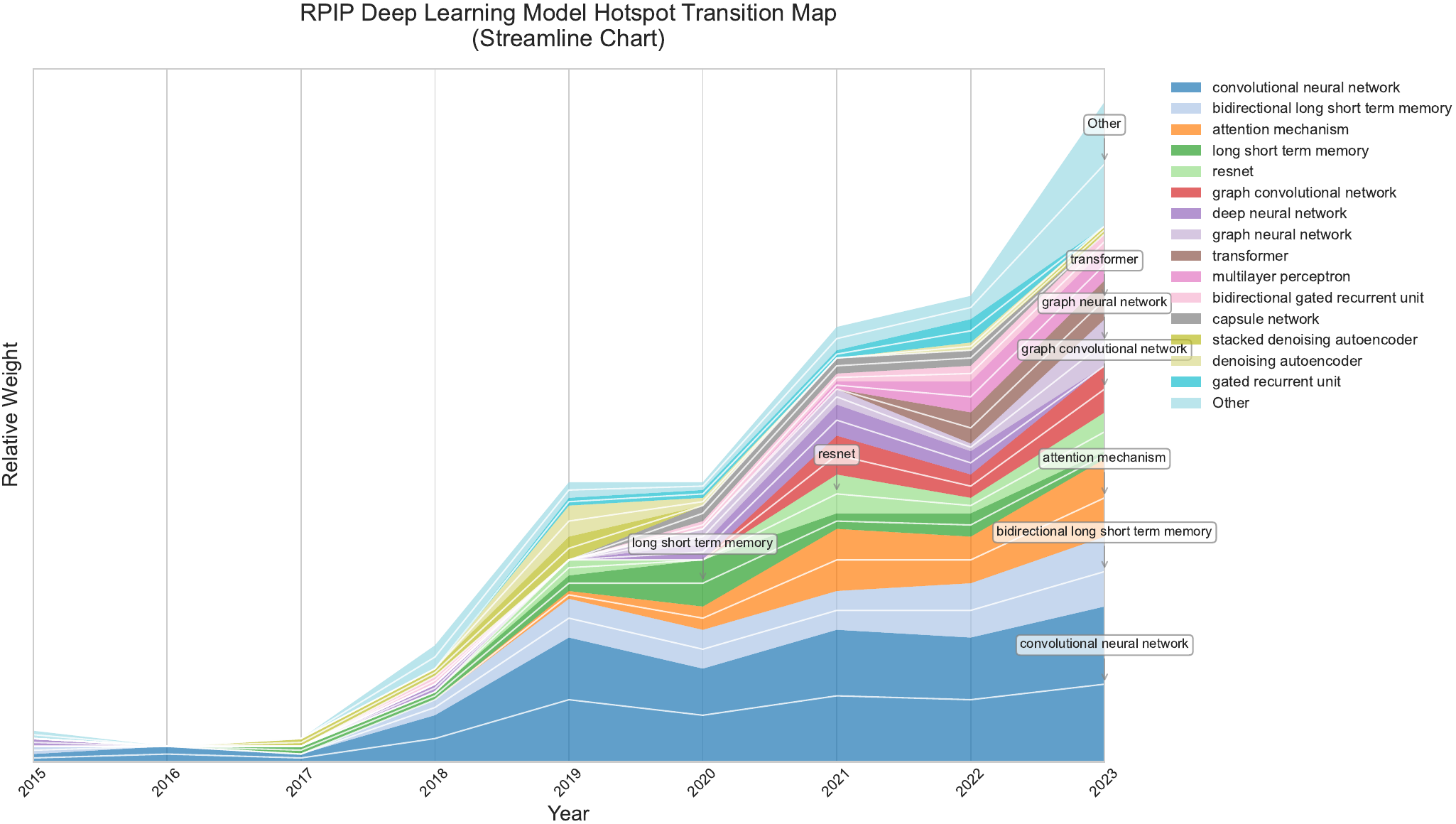}
    \caption{RPIP deep learning model hotspot transition map.}
    \label{FIG:hotspot_models}
\end{figure}

We categorized the models according to their backbone and classical variant models. 
We conducted separate statistical analyses for hybrid models, which incorporate more than one model type. 
Figure~\ref{FIG:hotspot_models} presents the hotspot transition map of DL models used in RPIP. 
As shown in the figure, 2018 was mainly dominated by early autoencoders, followed by the rise of CNN models in 2019. 
The peak period for RNNs was 2020, with research focus shifting to attention mechanisms in 2021. 
By 2022, GNNs had matured, and 
the hybrid models became mainstream in 2023.
We define models developed after 2021 as novel generation architectures, with GNNs and Transformers being representative paradigms.

\subsection{Basic models}
This section introduces basic DL models, exploring their innovations and variants.

\subsubsection{CNN-based models}

A CNN comprises convolutional layers (for local feature extraction), pooling layers, and fully connected layers.
Its hierarchical architecture enables progressive abstraction of higher-order features through local receptive fields, making it particularly suitable for RNA-protein motif-recognition tasks.
CNN-based models can process structured data and capture complex patterns through hierarchical feature extraction: 
This method is mainly used for extracting features from RNA and protein sequences or structures. 
Through the convolutional layers, CNN-based models can capture the local patterns in the sequences that may be important for predicting interactions~\cite{geraseva2023nucleic,yang2021rna}.
Wang et al.~\cite{wang2018combining} used CNNs to mine hidden high-level discriminative features from RNA and protein sequences.
Huang et al.~\cite{huang2022ncrna} designed a parallel CNN model that employs two one-dimensional CNNs to extract features from RNA and protein sequences, and uses a two-dimensional CNN to capture their cross-correlation information for RPIP.
Pan et al.~\cite{pan2022mcnn} proposed a method to predict binding sites by integrating multiple CNNs constructed from RNA sequence information extracted by windows of different lengths.
Ma et al.~\cite{ma2019concurrent} applied a CNN to identify RNA-binding residues on proteins.
In the prediction of RBP binding sites, Chung et al.~\cite{chung2019prediction} used a deep CNN model to process sequence and structural information. 
The design of multi-sized filters enables the simultaneous extraction of sequence motifs, structural motifs, and combined motifs.
Another important application of the convolutional layers is in motif detection. 
By extracting the parameters of the convolutional layer and converting them into a positional weight matrix, combined with the \textit{Multiple Expectation Maximization for Motif Elicitation} (MEME) tool, the binding sites between RNAs and proteins can be displayed~\cite{zhang2020idrbp_mmc,tayara2020improved}.

\textbf{\textit{Variants and Breakthroughs:}}
CNN-based models have strong feature-extraction ability, good computational efficiency, and strong adaptability in RPIP; but they also have disadvantages, such as global information limitations, large numbers of parameters, poor interpretability (they are often ``black boxes'', without access to the internal workings), and sensitivity to sequence length.
To address these issues, several variants based on CNN models have been proposed and applied to RPIP. 
Wang et al.~\cite{wang2021predicting} proposed an RPIP method that uses a pseudo-Siamese neural network constructed with \textit{Residual Network} (ResNet) architecture. 
This method simultaneously inputs the embedded features of both RNA and protein sequences into the network for end-to-end learning.
Zhao et al.~\cite{zhao2021multi} used ResNet to enhance feature-learning capabilities to infer protein-RNA binding preferences and predict novel interactions.

The capsule network~\cite{shen2019capsule,wang2021identifying} was proposed to address the drawbacks of CNNs (such as the loss of spatial information in pooling layers). 
It inherits the local receptive fields and weight-sharing characteristics of CNNs, but introduces a dynamic routing mechanism. 
Its use of inner product operations between vectors preserves the features' spatial relationships, and avoids information loss caused by pooling.  
Shen et al.~\cite{shen2019capsule} modified the original capsule-network code to simultaneously handle sequence and structural features for predicting RNA-protein binding preferences.
Li et al.~\cite{li2021capsule} proposed Capsule-LPI, a multi-channel capsule-network framework that integrates multi-modal features for RPI prediction:
Capsule-LPI is composed of four feature-learning subnetworks and one capsule subnetwork, and can capture the relationships between features.

\subsubsection{RNN-based models}

An RNN consists of recurrent units (for temporal hidden state propagation) and timestep unfolding (to model sequential dependencies).
As a memory-equipped architecture, its output is influenced by earlier inputs. 
The gating mechanisms in RNNs capture long-range dependencies, making them particularly suitable for feature extraction from RNA and protein sequences.
RNN-based models have been used more for processing sequential dependencies data.
In RPIP, they are used to capture long-term dependencies between RNA and protein sequences. 
Uhl et al.~\cite{uhl2021rnaprot} proposed an RNN-based model that performs binary classification on input sequences, and can predict protein-binding sites on RNA.
However, due to the vanishing/exploding gradient issues in traditional RNN architectures, their application in RPIP remains limited.

\textbf{\textit{Variants and Breakthroughs:}} 
Many variants have been proposed to address the problems of vanishing and exploding gradients, including:
\textit{Long Short-Term Memory} (LSTM)~\cite{wekesa2020lpi}, \textit{Gated Recurrent Unit} (GRU)~\cite{shen2019rna}, and \textit{Bidirectional Long Short-Term Memory} (BiLSTM)~\cite{wei2022deepstack}.

Wekesa et al.~\cite{wekesa2020lpi} proposed a DL method based on sequence features and a compact LSTM network:
It uses an RNN to learn discriminative features characterizing long-term dependencies between sequences, aiming to predict potential plant RPIs.
Yi et al.~\cite{yi2020unified} proposed \textit{Biological Sequence to Vector} (BioSeq2vec), a DL method for broad-spectrum biological sequence-representation based on the LSTM encoder-decoder model. 
BioSeq2vec can output effective fixed-length feature vectors for input nucleic acid or amino acid sequences of arbitrary lengths, which applies to RPI prediction.
Wei et al.~\cite{wei2022deepstack} used a stacked ensemble classifier composed of BiLSTM and GRU to predict RBPs.
These models use contextual information from sequence features to extract local patterns and long-term dependencies.

\subsubsection{GAN-based models}

A GAN consists of a generator (synthesizing artificial data) and a discriminator (distinguishing real versus generated samples).
Using both authentic and synthetic data, the generator progressively enhances its synthetic capability through adversarial training. 
This is particularly valuable for augmenting limited training datasets.
Although the direct application of GAN-based models in RPIP may be limited, the prediction-model performance can be enhanced by generating feature vectors. 
Similar to adversarial training, introducing noise into DL models to generate high-quality sample data can improve the feature learning and expression capabilities of DL models.
This can make it possible to explore and predict the RPI.
Wang et al.~\cite{wang2023rpi} applied the generative adversarial concept to RPIP to produce RPI-CapsuleGAN, which enhances the model's interpretability, feature learning, and expressive capabilities while effectively addressing the vanishing hierarchical structure issue in spatial modeling. 
Although GANs can partially mitigate data scarcity in RPIP, their performance depends on the quality of training data:
Biases or insufficient diversity in training data may compromise sample generation fidelity and optimization efficacy. 
Because of these things, pure GAN architectures have not yet become common in mainstream RPIP applications.

\subsubsection{DBN-based models}

A DBN employs stacked restricted Boltzmann machines trained layer-wise in an unsupervised manner. 
This pre-training strategy extracts hierarchical representations from biomolecular data (e.g., RNA/protein sequences or structures), enabling automated feature learning and dimensionality reduction. 
The multilayered architecture captures complex data patterns, yielding informative representations of RNAs and proteins.
Consequently, DBNs learn powerful multilevel features from inputs such as protein data. 
These learned representations serve as highly effective input for downstream tasks such as classifying RBPs or predicting molecular interactions~\cite{du2018deepmvf}.

Yang et al.~\cite{yang2018lncadeep} proposed a method that constructs a DBN to identify LncRNAs by stacking restricted Boltzmann machines. 
The method predicts RPIs using sequence and structural information.
Zhang et al.~\cite{zhang2016deep} used a DBN model to integrate feature maps derived from RNA primary sequences, predicted secondary structures, and tertiary structures into a unified representation, which was then used to identify RBP binding sites.
Latent hierarchical representations of RNAs and proteins
---
encompassing sequence motifs, structural domains, and interaction patterns
---
can be learned effectively by DBNs. 
Although DBNs have robust semi-supervised capabilities, the scarcity of high-confidence experimentally validated interactions fundamentally constrains their optimization in RPIP.

\subsubsection{DAE-based models}

A DAE comprises an encoder (for noise-robust compression) and a decoder (for data reconstruction), optimized through reconstruction-loss minimization.
This architecture enhances noise robustness in RPI data, enabling effective handling of low-quality experimental datasets.
DAE-based models are mainly used for feature-layer learning. 
This process helps the auto-encoder to learn effective low-dimensional data representations, and can also be used for various other tasks, such as data denoising and dimensionality reduction~\cite{pan2016ipminer}.
Due to the complex and diverse features of RNAs and proteins, auto-encoders can extract high-level abstract features that can facilitate subsequent RPIP~\cite{wekesa2019hybrid}.

Zhou et al.~\cite{zhou2019prediction} proposed a sequence-based method for predicting plant RPIs that uses stacked DAEs to automatically extract high-level features from RNA and protein sequences.
When predicting RPIs, Zhou et al.~\cite{zhou2021prpi} used stacked DAEs to extract sequence and structure features from LncRNAs and proteins.
However, these models had limited sensitivity to subtle variations in nucleotides or amino acids. 
DAE-based models may also fail to capture interaction-specific motifs, but may have robust performance when feature distributions diverge significantly. 
Consequently, in RPIP, DAEs typically serve as feature extractors or pretraining modules, rather than standalone predictors.

\subsubsection{MLP-based models}

An MLP has an input layer, one or more hidden layers, and an output layer featuring activation functions for non-linear transformations. 
As the most fundamental fully connected network, data propagates in one direction from the input layer through hidden layers to the output layer. 
It is commonly deployed as the terminal classifier in composite neural architectures.

MLPs can process and fuse various types of RNAs and protein features~\cite {wang2020matrix}.
MLP-based Models can extract feature representations and train feed-forward models with back-propagation~\cite{zhang2023iesslnc}.
To solve the problem of cross prediction, an enhanced regularization noise layer can be added to the MLP predictor to generalize the model.
Arican et al.~\cite{arican2023preddrbp} developed an MLP-based predictor for the multi-class classification of RBPs and other protein types.
Peng et al.~\cite{peng2021finding} developed a DL framework based on dual-network neural and MLP architectures to identify RPIs.
MLP-based models are relatively simple DL models that are easy to train and deploy. 
Due to the complexity of RNAs and protein features, MLP-based models rely heavily on pre-processed features.

\subsection{Novel models}

\subsubsection{GNN-based models}

\begin{figure}[!t]
    \centering
    \includegraphics[width=\textwidth]{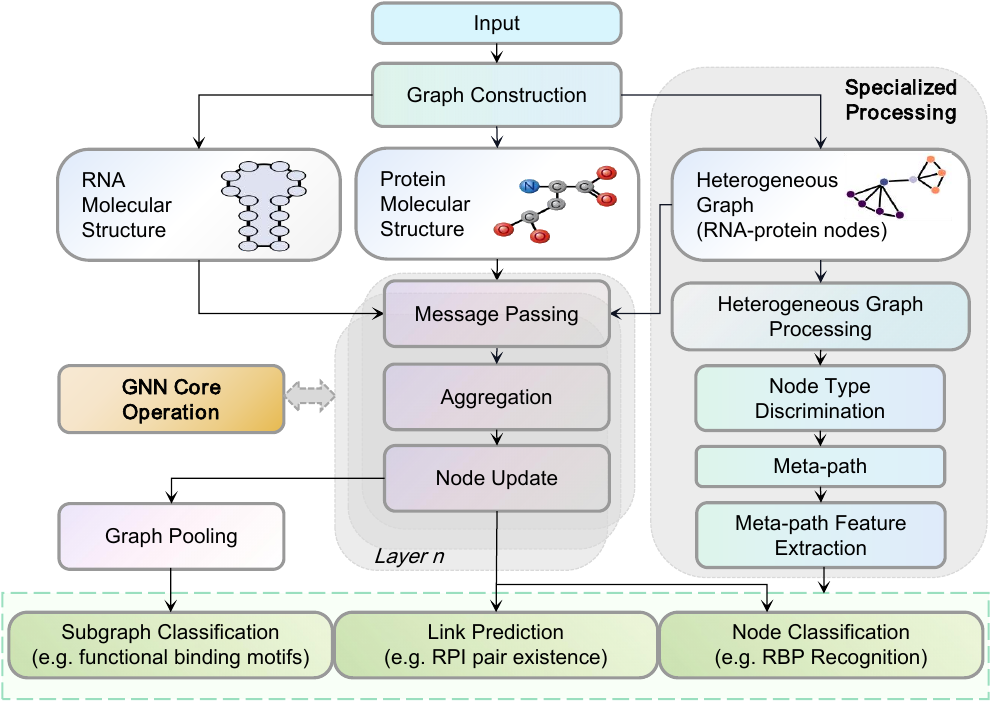}
    \caption{GNN-based models workflow diagram.}
    \label{FIG:GNN_Workflow_Diagram}
\end{figure}

The GNN operational mechanism involves three core stages: 
message passing, node updating, and graph pooling for output feature extraction.
This framework is particularly well-suited for modeling RPI networks, enabling the capture of topological characteristics at molecular binding interfaces, and other structural properties.
GNN methods can predict not only the links in the RNA-protein network (RPI pair existence), but also the subgraph classification (e.g., identifying functional binding motifs). 
Link-prediction and subgraph classification are the main applications of GNN models.
The RPIP problem is represented by an adjacency matrix~\cite{wei2023lpih2v,pv2020heterogeneous}, using deep encoders to provide better robustness for RPI sparse networks~\cite{zhao2022comparison}.
GNN-based models can also construct RNA and protein molecular structure graphs to capture structural features, which is very useful in RPIP.
Figure~\ref{FIG:GNN_Workflow_Diagram} shows the workflow of GNN-based models, covering core operations and specialized processing methods.

Arora et al.~\cite{arora2022novo} used two \textit{Graph Convolutional Network} (GCN) layers, taking advantage of the message-passing mechanism to learn node embeddings, and to capture the global network structure information of RNA and proteins (which is applied to predict RPIs).
Liu et al.~\cite{liu2021rna} proposed a GCN-based RNA secondary structure representation network for predicting binding sites. 
The network constructs a graph model from the RNA secondary structure to learn the topological properties of RNA.
GCN realizes the representation learning of graph nodes by aggregating the feature information of nodes and their neighbors. 
It updates node features through graph convolution operations, capturing the graph's local topological structure, and the dependency relationships among node features.  
\textit{Graph Sample and Aggregate} (GraphSAGE) was designed to address limitations of traditional GCN for handling dynamic graphs or unknown nodes: 
Instead of directly training embeddings for each node, it learns the aggregation function of node features, enabling generalization to unseen nodes.
Shen et al.~\cite{shen2021npi} used inductive node-representation learning with GraphSAGE, generating node embeddings through feature aggregation of neighboring nodes to predict RPIs.
Ma et al.~\cite{ma2023predicting} combined bipartite graph embedding with GNNs:
Although RPI heterogeneous graphs can be adapted through forced homogenization of node features, this process compromises critical molecular type discrimination.

\textbf{\textit{Variants and Breakthroughs:}}
GCN and GraphSAGE are more suitable for homogeneous graphs, but the RPI graph is heterogeneous: 
Some research, therefore, has focused on learning features from heterogeneous graphs (including those of RNA and protein nodes), as well as the interaction links between them.

\begin{figure}
    \centering
    \includegraphics[width=0.8\textwidth]{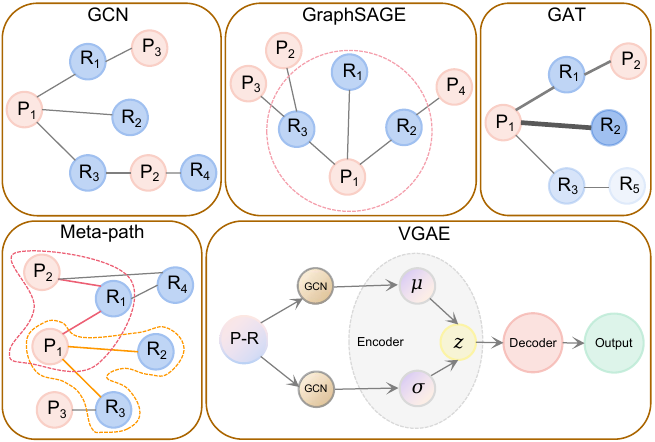}
    \caption{Schematic of common GNN-based models and techniques.}
    \label{FIG:GNN_models}
\end{figure}

In heterogeneous RPI graphs, meta-paths extract correlated features by defining semantic paths, enabling the prediction of RPIs.
A \textit{Graph Attention network} (GAT) can enhance feature representation in diverse sequences and structural descriptors~\cite{jiang2023structure}, efficiently learning the embedded features of target nodes (RNA protein pairs)~\cite{han2023ncrpi}, and improving prediction performance.
When RPI-labeled data are scarce or difficult to obtain, a \textit{Variational Graph Auto-Encoder} (VGAE) can learn potential node representations.
Figure~\ref{FIG:GNN_models} shows some commonly used GNN-based models and key techniques.
GCN~\cite{kipf2016semi} performs node representation learning for globally homogeneous graphs through spectral graph convolution.
GraphSAGE~\cite{hamilton2017inductive} uses sampling and aggregation mechanisms.
GAT~\cite{velivckovic2017graph} dynamically learns neighbor importance using attention weights.
VGAE~\cite{kipf2016variational} uses a variational autoencoder framework to learn latent variable distributions of graphs, performing link prediction and graph generation.
Meta-path~\cite{sun2011pathsim} is a semantic path template exclusive to heterogeneous graphs, guiding cross-type node-information propagation.

\subsubsection{Transformer-based models}

The Transformer~\cite{vaswani2017attention} comprises self-attention layers (for global dependency modeling), feed-forward layers, and residual connections with normalization.
By replacing recurrent operations with attention mechanisms, it can process global information in parallel while ignoring distance constraints. 
This enables the capture of salient features, and is very effective at predicting RNA-binding domains or functional sites in proteins.
Transformers can be used to model long-range dependencies between RNAs and protein features, and to capture complex RPI. 
The transformer's self-attention mechanism can capture the dependency relationships between all positions in the sequences, which is useful for predicting global interactions~\cite{wang2022self}.
Many of the networks listed above can be combined with transformer-based attention mechanisms to better capture features of RNAs and proteins~\cite{yang2023rbsp}.

Transformer-based biological language models have also been applied in RPIP. 
Like \textit{Bidirectional Encoder Representation from Transformers} (BERT) in natural language processing, DNABERT was pre-trained on large-scale datasets of DNA sequences to learn their deep semantic information.
Fine-tuning can be done by replacing \textit{Thymine} in DNA with \textit{Uracil} in RNA~\cite{du2022jlcrb,zhu2023dynamic}.
Protein sequences can be input into the attention-based 
Transformer architecture of ProteinBERT to predict RNA binding tendencies~\cite{jin2023hydra}.

\begin{figure}
    \centering
    \includegraphics[width=\textwidth]{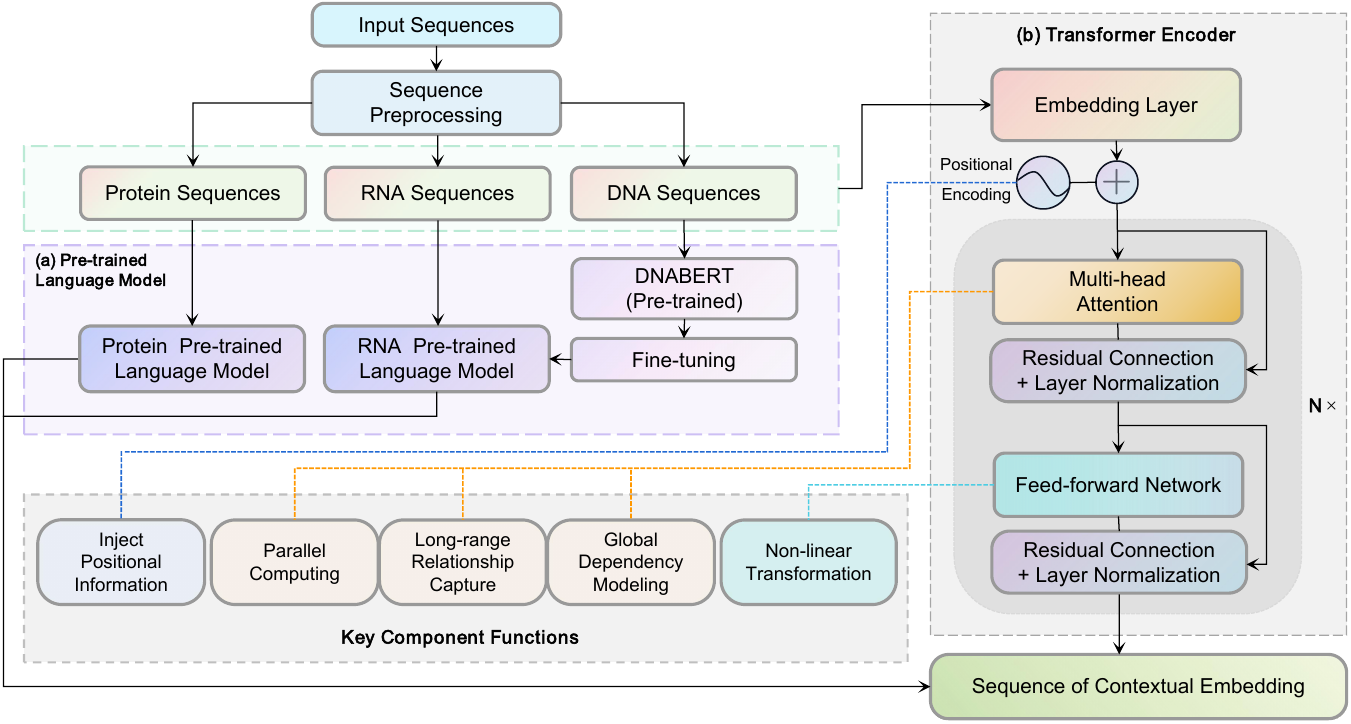}
    \caption{Transformer-based models workflow diagram. (a) Transformer-based pre-trained biological language models; (b) Models using the Transformer-encoder architecture.}
    \label{FIG:Transformer_Workflow_Diagram}
\end{figure}

Peng et al.~\cite{peng2022rbp} used the self-supervised pre-trained model \textit{Protein Text-to-Text Transfer Transformer-Xtra Large} (ProtT5-XL), which is based on the Transformer architecture, to generate embeddings of protein sequences for predicting RBPs.
Wang et al.~\cite{wang2022self} proposed a model to predict the binding sites between RNA and proteins:
This model is based on the Transformer's self-attention mechanism, combining positional encoding, multi-head attention, and position-wise feed-forward networks to extract sequence features.
Cao et al.~\cite{cao2023circssnn} developed a novel end-to-end circRNA binding site prediction model. 
They designed a network architecture named Seq-Transformer using Transformer to extract potential nucleotide dependencies.
Figure~\ref{FIG:Transformer_Workflow_Diagram} presents the workflow of Transformer-based models, covering both Transformer-based pre-trained biological language models, and models using the Transformer-encoder architecture.

\subsection{Hybrid models}
Hybrid DL models for RPIP are becoming a key technology for solving complex prediction problems. 
These models integrate multiple DL architectures, offering both diversity and flexibility. 
The complementary advantages of various model components can significantly improve prediction accuracy and robustness.
Hybrid models can, for example, integrate the advantages of CNNs' local perception, RNNs' contextual modeling, and attention's global association. 
From a structural perspective, hybrid models can be classified into three categories: 
cascading, co-processing, and hierarchical. 
\textit{Cascading} models use a linear pipeline such as CNN $\rightarrow$ BiLSTM, to process multimodal features (e.g., local features first, then global features).
\textit{Co-processing} models have parallel branches that independently process RNA/protein features and then concatenate the results (e.g., dual channels such as CNN$+$BiLSTM).
\textit{Hierarchical} models use deep interactive fusion (e.g., Transformer encoding sequence-structure interactions$+$graph attention for cross-modal fusion).

\subsubsection{Cascading models}
In cascading models, the output of one model serves as the input to another, forming a linear processing pipeline.
This architecture is often used to process the multi-modal features of RNA or proteins.
Liu et al.~\cite{liu2023crbsp} reported on a mid-term multimodal data fusion strategy for predicting circRNA-RBP binding sites. 
The model consists of two main branches: 
One uses a CNN to process the RNA sequence information, while the other uses a BiLSTM to analyze the secondary structure information. 
After the feature-extraction phase, the model concatenates the features derived from these two branches.
Niu et al.~\cite{niu2021rbpdl} extracted local conserved patterns of proteins with a CNN, combining sequence features and evolutionary features of proteins through a BiLSTM, and independently extracted structural patterns of proteins through another CNN.
Zhang et al.~\cite{zhang2019deepdrbp} proposed a two-layer predictor that shares the same input protein sequence:
It used a CNN to extract local features, and an LSTM to extract global features.

\subsubsection{Co-processing models}
Co-processing involves different model branches simultaneously and independently processing the same or different parts of the input, after which their outputs are fused.
Our survey of the literature revealed that, within co-processing models, both RNA and protein features mainly rely on sequence features. 
Very few studies have incorporated both sequence and structure features.
The independent branches of this architecture are often used to process RNA and protein features separately.
It is also suitable for the extraction of different features of the same biomolecule.
When predicting RPIs, Zhang et al.~\cite{zhang2018prediction} used a BiLSTM to capture long-range dependencies in RNA sequences, and a CNN to extract local patterns from protein sequences, with the two branches operating simultaneously and independently. 
Navamajiti et al.~\cite{navamajiti2019mcbel} constructed feature-processing branch structures for proteins and RNAs: 
A CNN and a BiLSTM were used to extract local features and sequence context information of the proteins, while a BiLSTM was used to extract sequence-dependent RNA features.
Feature concatenation was performed before the final prediction layer.
Wang et al.~\cite{wang2021predictingb} used CNNs to extract local features from RNA sequences and secondary structures, and BiLSTMs to capture global information and long-range dependencies:
Their model processes sequence and structural information in parallel before fusing the features for final prediction.
Pan et al.~\cite{pan2023wvdl} used CNNs to process RNA sequence features, BiLSTMs to handle RNA secondary structure features, and more CNNs to process annotation features.
Zhang et al.~\cite{zhang2022deeppn} used two independent processing channels:
A CNN was used to extract RNA sequence features, and a GCN was used to extract RNA secondary structure features. 
This model has been used to detect the binding sites of RBPs.

\subsubsection{Hierarchical models}
Hierarchical model components are fused at a deeper level, neither through simple input-output chaining nor final-output concatenation. 
Fusion occurs within internal model operations (e.g., replacing modules, introducing attention mechanisms, etc.).
Shen et al.~\cite{shen2023dare} proposed a novel DL model for predicting binding sites that consists of three hierarchical layers:
Layer 1 is a sequence-structure-aware Transformer encoder that
dynamically models the interaction between sequences and structures through multi-head self-attention mechanisms;
Layer 2 is a position-aware gated fusion unit that dynamically adjusts the contribution of sequence features and structure features; and 
Layer 3 is convolutional attention pooling that uses CNNs to extract local patterns from features, then aggregates global features through attention-based weighting. 
This final step further fuses local details and global contextual information to enhance prediction accuracy.
The RPI prediction model proposed by Wekesa et al.~\cite{wekesa2020deep} is also uses a three-layer architecture:
The bottom layer uses a LSTM to process RNA sequence features and a LSTM to process protein sequence features;
the middle layer uses graph attention to dynamically fuse cross-modal features; and
the top layer splices RNA-protein pairing features, and makes joint decisions.
GraphBind~\cite{xia2021graphbind} uses a multi-level graph representation for protein structures. 
GNN operations are applied to graphs at each level (such as the atom-, residue-, and surface-patch-level graphs). 
The inter-level information exchange mechanism is deeply integrated into the core message-passing framework, with multiple iterations of intra-layer and inter-layer information propagation.
For binding-site prediction, Yang et al.~\cite{yang2021icircrbp} used parallel branches of CNNs and bidirectional GRU for initial sequence and structure feature extraction, then fusing the sequence and structure features through a multi-level interactive attention network.

There has been increasing use of hybrid models, which can be very good at learning features and predicting. 
The combination of the good CNN local-feature extraction with the context-dependency extraction of LSTM or GRU can efficiently capture feature representation.
An attention mechanism can also be added to further enhance feature learning.
The node importance weights learned by GAT can be used as attention priors for LSTM networks, thereby guiding sequence modeling to focus on key functional regions.
The GCN can learn the topological RNA features through the secondary structure, and the CNN can then obtain the spatial importance of each RNA base, guiding the RNA secondary structure representation for RBP binding prediction.

\subsection{Model performance and limitations}

\setlength{\tabcolsep}{1pt} 
\begin{longtable}
    {>{\scriptsize}m{1.5cm} |
    >{\raggedright\arraybackslash \scriptsize}m{1.3cm} |
    >{\raggedright\arraybackslash \scriptsize}m{1.7cm} |
    >{\raggedright\arraybackslash \scriptsize}m{1.9cm} |
    >{\raggedright\arraybackslash \scriptsize}m{1.8cm} |
    >{\raggedright\arraybackslash \scriptsize}m{1.3cm} |
    >{\raggedright\arraybackslash \scriptsize}m{1.5cm} |
    >{\raggedright\arraybackslash \scriptsize}m{1.6cm}  } 
    \captionsetup{font=footnotesize}
    \caption{Multi-dimensional comparison of DL models for RPIP.}
    \label{tab:model_comparison} \\ 
    \hline
    \textbf{Models} & \textbf{Examples} & \textbf{Advantages} & \textbf{Limitations} & \textbf{Applicable tasks} & \textbf{Biological interpretability} & \textbf{Suitable datasets} & \textbf{Performance range (AUROC)}  \\ 
    \hline 
    \endfirsthead
    \caption[]{Continued.} \\ 
    \hline
    \textbf{Models} & \textbf{Examples} & \textbf{Advantages} & \textbf{Limitations} & \textbf{Applicable tasks} & \textbf{Biological interpretability} & \textbf{Suitable datasets} & \textbf{Performance range (AUROC)} \\ 
    \hline 
    \endhead
    \hline 
    \multicolumn{8}{r}{{\scriptsize Continued on next page.}} \\ 
    \endfoot
    \hline 
    \endlastfoot
    \multicolumn{8}{l}{%
    \begin{minipage}{\textwidth}
    \vspace{5pt}
    \scriptsize \textit{* Note:} The AUROC values in the table were derived from the results reported in the 179 articles collected. Since the AUROC in some articles differs across different datasets, we took the average. 
    AUROC values were not provided in some articles.
    \end{minipage}} 
\endlastfoot
 CNN-based & \cite{wang2018combining, huang2022ncrna, pan2022mcnn, ma2019concurrent, chung2019prediction, wang2021predicting, zhao2021multi, shen2019capsule, wang2021identifying, li2021capsule} & Strong local feature extraction, efficient motif recognition, computational efficiency & Ignores long-range dependencies, sensitive to sequence length, large parameters & Motif detection, binding site prediction, RPI binary prediction  & Variable &  RNA and protein sequence or structure datasets & $0.66-0.94$ \\ 
\hline
RNN-based  & \cite{uhl2021rnaprot, wekesa2020lpi, yi2020unified, wei2022deepstack} & Captures long-term dependencies in sequences & Limited long-term memory, computationally expensive & RPI binary prediction, RBP identification & Poor &  RNA or protein sequence datasets & $0.76-0.93$ \\ 
\hline
GAN-based & \cite{wang2023rpi}  & Enhances feature learning via synthetic data, improves model expressiveness & Limited direct application, hard to train, instability in adversarial training  & RPI prediction with data augmentation & Poor & Small scale RPI datasets  & $< 0.90$ \\ 
\hline
DBN-based &\cite{du2018deepmvf, yang2018lncadeep, zhang2016deep} & Learns complex patterns, effective for unsupervised pre-training & Computationally heavy, limited supervised learning capability & Unsupervised feature pre-training, binding site prediction & Poor & Sequence and structure datasets & $0.82-0.93$ \\ 
\hline
DAE-based & \cite{wekesa2019hybrid, zhou2019prediction, zhou2021prpi}  & Effective feature learning, denoising, dimensionality reduction & Requires careful tuning, limited direct applicability & Denoising, dimensionality reduction & Poor & Noisy or low-quality sequence data & $0.75-0.91$ \\ 
\hline
MLP-based  & \cite{arican2023preddrbp, peng2021finding} & Flexible feature fusion, fundamental DL architecture & Prone to overfitting, limited contextual awareness & Multi-class classification, hybrid model component & Poor & Structured feature vectors & $0.69-0.89$ \\ 
\hline
GNN-based & \cite{arora2022novo, shen2021npi, ma2023predicting, jiang2023structure, han2023ncrpi}  & Captures topological features, handles graph-structured data & Struggles with heterogeneous graphs, complex implementation & RPI binary prediction, binding site prediction, motif identification & Variable & Datasets representable as graphs (e.g.,
molecular structures, interaction networks) & $0.78-0.94$ \\ 
\hline
Transformer-based & \cite{jin2023hydra, peng2022rbp, wang2022self, cao2023circssnn}  & Models long-range dependencies, self-attention captures global interactions & Computationally intensive, high resource requirements & Binding site prediction, binding preference prediction, RBP identification & Moderate  & Large scale sequence data & $0.84-0.96$ \\ 
\hline
Cascading & \cite{liu2023crbsp, niu2021rbpdl, zhang2019deepdrbp}  & Combines strengths of multiple models & Linear pipeline may propagate errors, increased complexity & Multi-modal feature fusion, hierarchical feature extraction & Variable & Multi-source data & $0.82-0.91$ \\ 
\hline
Co-processing  & \cite{zhang2018prediction, navamajiti2019mcbel, wang2021predictingb, pan2023wvdl, zhang2022deeppn} & Parallel processing of diverse features & Requires feature alignment, fusion strategies critical for performance & Independent feature extraction, multi-source data integration & Variable & Multi-feature within single modality or cross-modality data &  $0.80-0.95$ \\ 
\hline
Hierarchical & \cite{shen2023dare, wekesa2020deep, xia2021graphbind, yang2021icircrbp} & Deep fusion, enhanced feature synergy  & Highly complex design, difficult to train and interpret & Cross-modal feature interaction, multi-level graph representation & Variable & High-dimensional structured data &  $0.85-0.96$ \\
\hline
\end{longtable}

Although RPIP models have made significant progress, they still face several challenges, which, as shown in Table~\ref{tab:model_comparison}, can be analyzed from several dimensions: 
models, advantages, limitations, applicable tasks, interpretability, applicable datasets, and performance range.
The performance ranges of all models depend on the specific dataset, task definition, data preprocessing pipeline, model architecture details, and hyperparameter tuning strategies. 
The ranges presented in the table are generalized summaries; substantial variations may occur in practical applications.

\begin{tcolorbox}[colback=white,sharp corners,boxrule=1pt]
    \textit{\textbf{Summary of answers to RQ4:}}
    \begin{itemize}
        \item[(1)] 
        In the field of RPIP, although basic models have made significant progress, they still have multiple limitations.
        Various variant models have attempted to make breakthroughs.
        
        \item[(2)]
        GNNs are an innovative tool for handling interaction networks. 
        They can directly model the topological structure of RNA-protein heterogeneous graphs (such as binding interfaces, molecular networks), breaking through the limitations of sequential models.
        
        \item[(3)] 
        Transformers achieve breakthroughs through their global attention mechanism. 
        The self-attention layer ignores distance constraints to capture full-sequence dependencies, addressing the challenge of long-range modeling faced by RNNs. 
        Pre-trained language models (such as DNABERT, ProtT5) transfer large-scale biological sequence knowledge, and improve the small-sample learning ability.
        
        \item[(4)] 
        Hybrid models have become the main solution, integrating multiple architectures.
        
        \item[(5)] 
        Model selection needs to balance task characteristics and data constraints. 
        CNNs/GCNs are suitable for identifying binding sites or binding motifs. 
        GNNs can be used for RPI network analysis. 
        The strength of Transformers/BiLSTMs lies in capturing long-range dependencies in RNA and protein sequences.
        
    \end{itemize}
\end{tcolorbox}

\section{Answer to RQ5: Validation and Evaluation Approaches for DL-based RPIP
\label{SEC:rq5}}

After the DL model is trained, its performance still needs to be evaluated.
This section introduces the model performance validation and evaluation metrics used in the literature.

\subsection{Evaluation metrics}

Common evaluation metrics for DL models include accuracy, precision, sensitivity (also known as recall), F1-score, and specificity.
These metrics are calculated using a confusion matrix by comparing model predictions with ground truth labels in the test set.
However, it is often difficult to obtain negative samples, which leads to imbalanced datasets:
In this case, the recall, precision, and F1-score are more appropriate measurements, because they can focus on the prediction of positive samples.
In addition, the AUROC can also be used to evaluate the classifier's performance.
The \textit{Matthews Correlation Coefficient} (MCC), and \textit{Recall-Precision Curve Area} (AUPR) are also good at measuring models' performance on imbalanced datasets.

\begin{table}
  \caption{Common evaluation metrics.}
  \label{tab:evaluation}
  \centering
  \begin{threeparttable}
  \scriptsize
  \renewcommand{\arraystretch}{1.4} 
     \setlength{\tabcolsep}{1mm}{
          \begin{tabular}{m{1.5cm}|c|m{2.5cm}|m{3.5cm}}
          \hline
  \textbf{Evaluation metrics} & \textbf{Formula} & \textbf{Description}& \textbf{Characteristics} 
  \\\hline
   Accuracy  & $ \frac{TP + TN}{TP + FP + TN + FN}$ 
   & Evaluate the overall classification accuracy of the model 
   & Greatly affected by the imbalance between RPIs and non-RPIs in the samples 
   \\\hline
   Sensitivity  & $\frac{TP}{TP + FN}$ & Indicates the model's ability to identify positive samples  & Can reduce the missed detection of valuable RPIs  \\\hline
   Specificity  & $\frac{TN}{TN + FP}$ & Indicates the model's ability to identify negative samples  & The ability to avoid misclassifying non-RPIs as RPIs, ensuring the reliability of the predicted RPIs \\\hline
   Precision & $\frac{TP}{TP + FP}$ & Assesses the reliability of the model's positive predictions  & Focuses on the proportion of actually true RPIs among the predicted RPIs \\\hline
   F1-score & $2 \times \frac{\text{precision} \times \text{recall}}
   {\text{precision} + \text{recall}}$ & Assesses the trade-off between precision and recall in positive sample detection  & Takes into account both ''finding more true RPIs'' and ''ensuring the reliability of predicted RPIs''  \\\hline
   MCC &  $\frac{TP\times TN-FP\times FN}{\sqrt{(TP+FP)(TP+FN)(TN+FP)(TN+FN)}}$ & Incorporates correlation coefficients of TP, TN, FP, and FN  & Reflects the model's ability to identify RPIs when there is class imbalance  \\\hline
   AUROC  & $\int\limits_{0}^{1} TPR(FPR) \, d(FPR)$ & Reflects the model's ability to discriminate between positive and negative samples  & Can be used to compare the performance of different RPI prediction models, or to evaluate the generalization ability of the same model under different datasets \\\hline
   AURP  & $\int\limits_{0}^{1} precision(recall) \, d(recall)$ & Commonly used to evaluate the model's comprehensive performance in identifying positive samples &  When there is class imbalance in RPI data and more attention is paid to the recognition performance of RPIs \\\hline
   Pearson correlation coefficient & $r = \frac{\sum_{i=1}^{n}(x_i - \bar{x})(y_i - \bar{y})}{\sqrt{\sum_{i=1}^{n}(x_i - \bar{x})^2} \sqrt{\sum_{i=1}^{n}(y_i - \bar{y})^2}} $ & The Pearson correlation coefficient measures the linear dependence between two continuous variables &  It can be used to evaluate the binding affinity or binding strength between RNA and proteins, etc. \\\hline
   MSE  & $ \frac{1}{n} \sum_{i=1}^{n}(\hat{y}_i - y_i)^2$ & MSE measures the degree of deviation between predicted and true values &  It can be used to evaluate the binding affinity or binding strength between RNA and proteins, etc. \\\hline
  \end{tabular}  
  }  
  \begin{tablenotes}
    \item[*] Note:where $TP$, $FP$, $TN$, and $FN$ refer to the number of \textit{True Positives}, \textit{False Positives}, \textit{True Negatives}, and \textit{False Negatives}, respectively; while
$TPR$ is the \textit{True Positive Rate}, which is the same as sensitivity, and $FPR$ is the \textit{False Positive Rate}, defined as 
$FPR = \frac{FP}{FP + TN}$
$X$ and $Y$ are two variables, $n$ is sample size,
$\bar{x}$ and $\bar{y}$ are the sample means.
$\hat{y}_i$ is predicted values,  and $y_i $ is true values.
        \end{tablenotes}
\end{threeparttable}
\end{table}

If the model is used for regression tasks~\cite{liu2022inferring}, validation of the in-vivo dataset can be done using measures such as the Pearson correlation coefficient, and the maximum \textit{Mean Square Error} (MSE)~\cite{du2022deep}.
Pearson correlation coefficient and MSE are often used to measure the binding affinity or binding strength between RNA and proteins.
Table~\ref{tab:evaluation} presents the main model-performance evaluation metrics used in DL-based RPIP.

The reliability of these metrics can be evaluated through the following approaches:
Repeated dataset splitting can be used to calculate metric-fluctuation ranges (e.g., standard deviation), showing the stability of the results;
hypothesis testing (e.g., t-test) can compare metrics across different models, showing whether or not differences are statistically significant; and
benchmarking against baseline or state-of-the-art models can show whether or not metric improvements have practical significance.

High-precision DL predictions require the adjustment and optimization of key parameters and factors across multiple stages and elements, as follows: 
\begin{itemize}
    \item Model architecture: depth (number of layers), width (number of units), activation-function selection, and regularization techniques (e.g., $L1$/$L2$ weight decay, dropout rate). 
    \item Data preparation: data quality and scale, preprocessing methods, and dataset-splitting ratios. 
    \item Training process: loss function design, optimizer selection, learning rate and scheduling strategies, batch size, number of training epochs, and gradient clipping threshold. 
    \item Evaluation and prediction: metric selection, validation strategy, and prediction-threshold setting (particularly in classification tasks).
\end{itemize}

The optimal combination of parameters depends on the specific task and data characteristics: 
Typically, the learning rate, model complexity (including numbers of layers/units), regularization strength, and data quality have the most significant impact on the model performance.

\subsection{Model validation}

A rigorous validation strategy is crucial for evaluating model performance. 
This section introduces several commonly used validation methods: 
\textit{Cross-validation} assesses model stability; 
\textit{cross-database validation} mitigates bias in dataset selection; 
\textit{cross-species validation} reveals limitations in generalization across species; \textit{in vitro and in vivo validation} bridges the gap between computational predictions and physiological contexts.

\textbf{\textit{Cross-validation:}} 
Cross validation is a method for evaluating model performance. 
It involves splitting the dataset into training, validation, and testing subsets.
5-fold and 10-fold cross validation are the most common. 
Leaving-one-out cross-validation~\cite{zhang2023iesslnc} is a special form of cross-validation.  

\textbf{\textit{Cross-database validation:}} 
Due to the particular characteristics of RPIP, some studies have trained models with one database, and retrained and validated with another~\cite{feng2022idrbp,wang2022self,yang2023rbsp}. 
This can measure the generalizability and robustness of models. 
However, because there may be overlapping datasets in different databases, it is necessary to carefully check for this in the validation database.

\textbf{\textit{Cross-species validation:}}
Cross-species validation has also been conducted, including training models for Homo sapiens and Escherichia coli, which were then fine-tuned on a trained Salmonella model~\cite{peng2022rbp}. 
Zhao et al.~\cite{zhao2022predicting} trained models on the dicots Glycine max and monocots Zea mays datasets, and validated on the dicots Arabidopsis thaliana dataset.
This validation method requires that the models have high generalizability and robustness.
There have not yet been many studies on cross-species validation.
The model proposed by Wang et al.~\cite{wang2018combining}, for example, performs well for eukaryotic model organisms, but its accuracy significantly decreases (by approximately 15–17\%) when applied to Escherichia coli (a prokaryote). 
The method developed by Jin et al.~\cite{jin2023hydra} achieved an AUROC of 0.895 and an AUPR of 0.745 on human datasets. 
During cross-species validation, AUROC dropped by about 10\%.  
Wang et al.~\cite{wang2023rpi} used independent validation datasets from six species to assess model generalization. 
They found significant performance degradation, with a 13.54\% decline in Escherichia coli, indicating weak generalization to distantly related species. 
Even in humans, the accuracy dropped by 9.04\%, highlighting challenges in cross-species prediction within eukaryotes. 
When a human-trained model by Zhang et al.~\cite{zhang2021predicting} was directly applied to mouse data, accuracy fell to 0.517 (a 26.7\% decrease). 
Zhang et al. \cite{zhang2022prerbp} confirmed that evolutionary distance substantially impacts cross-species prediction, with AUPR losses averaging 7–10\% within taxonomic groups and reaching 20–25\% across groups. 
These findings highlight species-dependent limitations in model generalization.

Cross-species validation is an important methodology for evaluating model-generalization capabilities. 
The enhancement of cross-species validation efficacy can take advantage of recent advances in transfer learning and neural network optimization techniques:
For instance, Khatir et al.'s meta-heuristic algorithm for neural network optimization \cite{khatir2025enhancing}, integrated with a frequency-response feature-enhancement strategy, significantly improves model performance across multiple biological contexts. 
Their proposed multimodal feature fusion framework~\cite{khatir2024advancing} further enhances generalization capability on heterogeneous datasets through adaptive learning mechanisms. 
This has great potential for addressing critical challenges in DL-based RPIP during cross-species validation.

\textbf{\textit{In vitro and in vivo validation:}}
The representative {\em in vivo} RPI is \textit{Deep Learning for Protein-RNA Binding} (DLPRB)~\cite{ben2018deep}, which uses both a CNN and an RNN.
The representative {\em in vitro} RPI is \textit{Thermodynamic Prediction} (ThermoNet) \cite{su2019integrating}, which uses a CNN for RPIP. 
Most models are trained and validated on {\em in vitro}  experimental datasets, with very few combining {\em in vitro} and {\em in vivo}~\cite{du2022deep,liu2022inferring}.
This type of validation typically involves two tasks: {\em in vivo} binding-strength prediction (a regression task) and {\em in vivo} binding-site prediction (a classification task).
These tasks are often used to construct a deep neural network model for multi-task learning. 
The Pearson correlation coefficient is often used to measure the predicted binding strength between the RBP and a given RNA in {\em in vitro} validation. 
{\em In vivo} validation uses AUROC to predict whether or not the RNA binds to proteins~\cite{du2022deep,karin2021multirbp,su2019integrating}.

Table~\ref{tab:In_vitro_In_vivo} presents a comparison of typical protein-RNA binding predictions in {\em in vivo} and {\em in vitro} scenarios. 
In practical research, amongst other things, {\em in vivo} studies may also involve binding-strength analysis, and {\em in vitro} experiments may also explore binding sites. 

\begin{table}[!t]
    \caption{In vitro vs. in vivo comparison.}
    \label{tab:In_vitro_In_vivo}
    \centering
    \begin{threeparttable}
        \scriptsize
        \renewcommand{\arraystretch}{1.4} 
        \setlength{\tabcolsep}{1mm}{
        \begin{tabular}{m{3cm}|m{4.8cm}|m{4.8cm}}
            \hline
            \textbf{Comparison dimension} & \textbf{In vitro prediction} & \textbf{In vivo prediction}\\\hline
            Prediction-task type  &  Regression problem & Classification problem
            \\\hline
            RPI type   & Binding-strength prediction & Binding-site presence prediction
            \\\hline
            Evaluation metrics   & Pearson correlation coefficient  & AUROC
            \\\hline
            Advantages of multi-task learning   & Learning general binding patterns, enhanced generalization & Cross-task knowledge transfer
            \\\hline
            Challenges   & Difficulty simulating cellular environment  & Significant environmental noise, degraded generalization performance
            \\\hline
        \end{tabular}}  
    \end{threeparttable}
\end{table}

\begin{tcolorbox}[colback=white,sharp corners,boxrule=1pt]
  \textit{\textbf{Summary of answers to RQ5:}}
    \begin{itemize}
        \item[(1)] 
        There are many model-validation methods. 
        Cross-validation evaluates model stability by dividing datasets; 
        cross-database validation tests model generalizability, but attention should be paid to the overlap of datasets in different databases; 
        cross-species validation enhances the adaptability of models in heterogeneous biological environments;
        most models rely on in vitro data, with only a few combining with in vivo validation.
        In-vitro and in-vivo validation have clinical translatability, but there is a gap between their applicability.
        
        \item[(2)]
        The selection of evaluation metrics depends on data characteristics and task types. 
        Accuracy is commonly used for balanced datasets; but
        other metrics are more suitable for imbalanced datasets (precision, recall, F1-score, AUROC, MCC, AUPR, etc.); 
        regression tasks (such as in vivo dataset validation) use indicators like Pearson correlation coefficient and MSE to quantify the binding strength of the RNA-protein combination.
        
        \item[(3)] The surveyed literature indicates that some evaluation metrics have been widely recognized and applied.   
    \end{itemize}
\end{tcolorbox}

\section{Answer to RQ6: Application Domains of DL-based RPIP
\label{SEC:rq6}}

DL models generate interaction probability scores via \textit{sigmoid} or \textit{softmax} output layers, where higher values indicate greater interaction likelihood. 
Thresholds are optimized by maximizing validation metrics (e.g., F1-score) or customizing loss functions. 
The validated model is then deployed for predictive applications.
DL-based RPIP has been applied to many different domains, as described in the surveyed literature.
This section summarizes these applications, which we categorize into the following groups.

\subsection{Novel association prediction}

\textit{$\bullet$ Finding associated proteins for new RNAs:}
By blocking the proteins associated with LncRNA:FGD5-AS1, the interacting proteins can still be predicted~\cite{peng2021finding}.
Similarly, hiding the interaction information for RNA:SNHG1 causes the model is try to discover proteins~\cite{peng2022enanndeep}. 
Deep gradient boosting decision trees can be used to identify the proteins interacting in LncRNA:RN7SL1~\cite{zhou2021lpi}.
A \textit{Hybrid framework} can also predict new RNA:RNase MRP through RPI~\cite{zhou2021lpib}.

\textit{$\bullet$ Finding associated RNAs for new proteins:}
When the information related to protein Q9H9G7 was hidden, \textit{Deep Learning framework with Dual-net Neural}  predicted the interacting RNAs~\cite{peng2021finding}.
With Q9UKV8 as a new protein, an \textit{Ensemble Learning Framework}~\cite{peng2022enanndeep} can be used to find its related LncRNAs.

\subsection{New RPI discovery}

Models have also been used to detect new RPIs~\cite{wang2018combining}.
\textcolor{blue}{}
Similar RNAs may interact with similar proteins, and some models can predict new RPIs using known ones~\cite{li2021lpi}.
The \textit{ab-initio method} model can predict RNA binding residues in proteins, which can enhance our understanding of RPI~\cite{liu2021aprbind}.
The \textit{Multi-scale Characterizing Sequence and Structure} model can not only perform RPIP, but can also predict multiple RNA binding sites \cite{zhang2023crmss}.

\subsection{Potential RBP-binding motif discovery}

The \textit{Dual-Aware Encoder} model extracts binding motifs for interaction sites \cite{shen2023dare}.
Zhang et al.~\cite{zhang2016deep} conducted experimental research into the recognition of binding sites. 
Their model can identify potential motifs, and explore RNAs binding to RBPs.
DeepCLIP has also undergone extensive validation on binding motifs~\cite{gronning2020deepclip}.

\subsection{Integrated molecular-function prediction}

\textit{$\bullet$ RNA function prediction:}
Kumar et al.~\cite{pv2020heterogeneous} examined the functional association prediction of LncRNAs in clinical practice.
Tan et al.~\cite{tan2021novel} demonstrated how using RPIP could predict LncRNA functions.
In the Zea mays datasets, some genes lack annotations, and their functions are unknown. 
Therefore, identifying their protein interaction
partners is an alternative way of annotating them.
Wekesa et al.~\cite{wekesa2019hybrid} explored RPIP in Arabidopsis thaliana and Zea mays (which are available online\footnote{\url{http://bis.zju.edu.cn/PlncRNADB}.}), subsequently annotating potential LncRNA functions.

\textit{$\bullet$ Protein function prediction:}
RPI not only affects the RNA function, but also affects the function of proteins. 
New RBPs involved in RNA biosynthesis can be identified through RPIP~\cite{zhang2019crip}.
An innovative deep multi-task architecture has been developed to predict the function and interaction regions of the protein:Sir3p from budding yeast~\cite{zhang2022deepdisobind}.
Transfer learning has been used to identify RBPs of two species with limited data, and has the potential to reveal new species~\cite{zhang2022prerbp}.

\subsection{Clinical application status and translation challenges}

This section critically evaluates the clinical utility and scientific significance of RPIP.

\textit{$\bullet$ Related case studies:}
DL for RPIP has advanced rapidly. 
While its direct adoption as a core tool for clinical diagnosis or treatment decision-making remains limited (particularly as primary diagnostic), it has already helped to bridge fundamental biological research and clinical medicine. 
DL is now delivering tangible impact across multiple critical clinical domains including virology~\cite{padhi2023transforming}, oncology~\cite{shimizu2020artificial}, neuroscience~\cite{richards2019deep}, and drug development~\cite{wu2024deep}.
DL-based RPIP has successfully identified some proteins associated with cancers~\cite{du2022deep}.
Wang and Lei~\cite{wang2022web} predicted proteins related to bladder cancer. 
Park et al.~\cite{park2021genome} investigated the risk areas of schizophrenia through RPI dysregulation.
A novel model for predicting RNA-binding sites based on protein-surface topography has been used to predict the RNA binding sites on the Ebola virus matrix for protein:VP40 (7K5L), and for the \textit{Severe Acute Respiratory Syndrome Coronavirus 2} protein:NSP16 \cite{li2022pst}.
LncRNAs activate diseases such as bladder tumors, cervical tumors, and thyroid cancer by interacting with the proteins~\cite{ma2023predicting}.

\textit{$\bullet$ Clinical-translatability analysis:}
Although DL has demonstrated theoretical breakthrough potential in RPIP, and has technical advantages in areas like as precision medicine and accelerated drug development, it still faces obstacles for clinical translation.
\begin{itemize}
        \item[(1)] Physiological complexity:
        Currently, almost all DL models are based on static binding prediction:
        Clinical requirements require understanding of dynamic metabolic processes.        
        \item[(2)] Tensions between data-driven approaches and causal mechanisms:
        Most DL-based RPIP models use RNA and protein features at the sequence or structural level. 
        However, clinical demands require understanding subcellular compartment-specific interactions. 
        Current DL models remain incapable of modeling physiological barriers such as nucleocytoplasmic transport.        
        \item[(3)] Discrepancies in validation:
        DL models often use AUROC for performance evaluation on validation datasets. 
        However, in RPIP, if a model predicts an RNA-binding protein as a cancer target, a deeper interpretation of its biological pathways is required.
\end{itemize}

\begin{tcolorbox}[colback=white,sharp corners,boxrule=1pt]
    \textit{\textbf{Summary of answers to RQ6:}}
    \begin{itemize}
        \item[(1)] 
        DL enables the prediction of RNA-protein molecular interactions, including:
        association prediction from RNA to protein, or vice versa;
        discovery of novel interaction pairs; and
        functional annotation of RBPs or RNAs.
        
        \item[(2)] 
        The studies surveyed in this paper provide not only a deeper understanding of the functions of RNAs and proteins, but also provide new ideas and methods for disease treatment and drug development.

        \item[(3)] 
        Clinical translation of DL-based RPIP has great potential, but also faces significant challenges.
    \end{itemize}
\end{tcolorbox}

\section{Answer to RQ7: Tools for RPIP
\label{SEC:rq7}}

\setlength{\tabcolsep}{0.8pt} 
\begin{longtable}
    {>{\scriptsize}m{1.8cm} |
    >{\raggedright\arraybackslash \scriptsize}m{1.5cm} |
    >{\raggedright\arraybackslash \scriptsize}m{2cm} |
    >{\raggedright\arraybackslash \scriptsize}m{1cm} |
    >{\raggedright\arraybackslash \scriptsize}m{1cm} |
    >{\raggedright\arraybackslash \scriptsize}m{1.5cm} |
    >{\raggedright\arraybackslash \scriptsize}m{1.3cm} |
    >{\raggedright\arraybackslash \scriptsize}m{2cm} |
    >{\raggedright\arraybackslash \scriptsize}m{1.1cm} } 
    \captionsetup{font=footnotesize}
    \caption{RPIP tools. 
    (\textbf{1 Star}: Completely white-box, all parameters, logic, and calculation processes are visible and traceable.
    \textbf{2 Stars}: Explainable model, the core logic is understandable to humans, but requires tool-assisted analysis.
    \textbf{3 Stars}: Grey Box, the overall structure is transparent, but contains non-linear/complex modules internally.
    \textbf{4 Stars}: Dark Box, decision paths are difficult to trace, relying on post-hoc explanation tools.
    \textbf{5 Stars}: Completely black-box, internal mechanisms are fully closed to humans.)}
    \label{tab:tools_analysis} \\ 
    \hline
    \textbf{Name} & \textbf{URL} & \textbf{Brief introduction} & \textbf{Input formats} & \textbf{Output format} & \textbf{Output information} & \textbf{Scalability} & \textbf{Limitations} & \textbf{Black-box predictions} \\ 
    \hline 
    \endfirsthead
    \caption[]{Continued.} \\ 
    \hline
    \textbf{Name} & \textbf{URL} & \textbf{Brief introduction} & \textbf{Input formats} & \textbf{Output format} & \textbf{Output information} & \textbf{Scalability} & \textbf{Limitations} & \textbf{Black-box predictions}  \\ 
    \hline 
    \endhead
    \hline 
    \multicolumn{9}{r}{{\scriptsize Continued on next page.}} \\ 
    \endfoot
    \hline 
    \endlastfoot
    CRWS \cite{wang2022web} & \url{http://www.bioinformatics.team}  & It provides the prediction for RBP binding sites on circRNA 
    & FASTA (RNA), RBP name
    & Online results, PNG 
    & Binding probability, potential binding sites
    & Web server
    & Fixed-length sequences, dependency on training data sets
    &\ding{72}\ding{72}\ding{72}\ding{72}\ding{73}
    \\\hline
    
    iDRBP-ECHF ~\cite{feng2022idrbp} & \url{http://bliulab.net/iDRBP-ECHF} & It provides the prediction for DNA-binding proteins and RNA-binding proteins 
    & FASTA (protein), TXT (protein) 
    & TXT, PNG
    & Binding probability, binary prediction labels
    & Web server
    & Lengthy computation time, dependency on training data sets, limited applicability to novel proteins
    &\ding{72}\ding{72}\ding{72}\ding{72}\ding{73} 
    \\\hline
    
    PredDRBP-MLP ~\cite{arican2023preddrbp} & \url{https://sourceforge.net/projects/preddrbp-mlp} & It provides the prediction for DNA binding protein and RBP, as well as non-nucleic acid binding proteins
    & FASTA (protein) 
    & Results displayed in the local terminal
    & Probability scores 
    & Local version
    & Outdated dependencies and obsolete model
    &\ding{72}\ding{72}\ding{72}\ding{72}\ding{73}
    \\\hline
    
    ProNA2020 ~\cite{qiu2020prona2020} & \url{https://predictprotein.org/} & It provides predictions of binding sites and binding residues for protein interactions with DNA, RNA, and other proteins
    & FASTA (protein)
    & Result online, PDB, TXT
    & Functional sites, residue-level prediction
    & Web server
    & Sequence length limit, long computation time, limited batch processing support
    &\ding{72}\ding{72}\ding{72}\ding{72}\ding{73} 
    \\\hline
      
    NucleicNet ~\cite{li2019deep}  & \url{http://www.cbrc.kaust.edu.sa/NucleicNet/} &It provides the prediction for binding preference of RNA components at any position on a given protein surface
    & PDB 
    & CSV, HTML, PNG
    & Residue binding probability, binding preference score
    & Web server, API interface, Docker image provided
    & Cannot predict binding sites in intrinsically disordered regions.Cannot distinguish specific binding modes
    &\ding{72}\ding{72}\ding{72}\ding{72}\ding{73}
    \\\hline
    
    IPMiner ~\cite{pan2016ipminer} & \url{http://www.csbio.sjtu.edu.cn/bioinf/IPMiner} & It provides the prediction for interaction between ncRNA and proteins
    & FASTA (RNA and protein)
    & TSV, TXT, CSV, PNG
    & Interaction score
    & Download package provided.
    & Only binary interactions supported. Limited species coverage
    &\ding{72}\ding{72}\ding{72}\ding{72}\ding{73}
    \\\hline
    
    PrismNet-variants ~\cite{sun2021predicting} & \url{http://prismnet.zhanglab.net/} & It provides prediction for dynamic cellular protein-RNA interactions 
    & FASTA (RNA), protein name
    & TSV, PNG
    & RBP binding sites, RBP binding probability.
    & Command-line tool provided. Pre-trained models provided
    & Limited to pre-trained RBPs, incomplete RBP coverage, species-specific
    &\ding{72}\ding{72}\ding{72}\ding{72}\ding{73}
    \\\hline
    
    DeepBind ~\cite{alipanahi2015predicting} & \url{http://tools.genes.toronto.edu/deepbind/} & It is an online repository of 927 DeepBind models representing 538 distinct transcription factors and 194 distinct RBPs  
    & FASTA (RNA / DNA)
    & CSV, TSV 
    & Binding affinity score
    & Web server, local version
    & Covers only pre-trained protein/RNA models
    &\ding{72}\ding{72}\ding{72}\ding{72}\ding{73}
    \\\hline
    
    RBPNet ~\cite{horlacher2023towards} & \url{https://biolib.com/mhorlacher/} & It provides pre-trained models of RNA sequences and prediction for signal spectra and sequence attribution maps
    & FASTA (RNA)
    & TSV, TXT 
    & Binding probability
    & Provided via Docker containers
    & Fixed input length, species limitation, model dependency
    &\ding{72}\ding{72}\ding{72}\ding{72}\ding{72}
    \\\hline
      
    RBPsuite ~\cite{pan2020rbpsuite} & \url{http://www.csbio.sjtu.edu.cn/bioinf/RBPsuite/} & It provides the prediction for the binding site of multiple species of RNAs (linear RNAs and circular RNAs) 
    & FASTA (RNA), RNA sequence
    & CSV, TSV, TXT
    & Binding score, binding sites
    & Web server, no API support
    & Only supports built-in RBPs
    &\ding{72}\ding{72}\ding{72}\ding{72}\ding{73}
    \\\hline
    
    DeepCLIP ~\cite{gronning2020deepclip} & \url{http://deepclip.compbio.sdu.dk} & It provides binding preferences of RBPs
    & RNA sequence, protein sequence 
    & Result online 
    & Binding probability 
    & Web server
    & RNA length limitation, lacks batch processing, training data bias
    &\ding{72}\ding{72}\ding{72}\ding{73
    }\ding{73}
    \\\hline
    
    DeepDISOBind ~\cite{zhang2022deepdisobind} & \url{https://www.csuligroup.com/DeepDISOBind/} 
    & It provides the prediction for the disordered residues that interact with proteins, DNA, and RNA
    & FASTA (protein)
    & CSV, TSV, PNG 
    & RNA/DNA-binding propensity
    & Web server, API support
    & Sequence length limit, bias in training dataset, function prediction restricted to disordered regions
    &\ding{72}\ding{72}\ding{72}\ding{72}\ding{73}
    \\\hline
    
    PrismNet ~\cite{xu2023prismnet} & \url{http://prismnetweb.zhanglab.net} & It provides the prediction for RBP binding sites in different cellular contexts
    & RNA sequence, RNA structure, TXT, TSV
    & Result online, JPG, TXT, PNG 
    & Probability score, binding site
    & Web server, API access requires approval
    & Sequence length limits, support restricted to 7 modifications, inherent training data bias
    &\ding{72}\ding{72}\ding{72}\ding{72}\ding{73}
    \\\hline
    
    PST-PRNA ~\cite{li2022pst} & \url{http://www.zpliulab.cn/PSTPRNA} 
    &  It provides the prediction for RNA binding sites on proteins
    & PDB
    & Result online, CSV
    & Binding probability
    & Web server
    & Species limitation, surface feature computation is computationally intensive
    &\ding{72}\ding{72}\ding{72}\ding{72}\ding{73}
    \\\hline
    
    iDeepMV ~\cite{yang2021rna}  & \url{http://www.csbio.sjtu.edu.cn/bioinf/iDeepMV} & It provides the prediction for which RBPs an unexplored RNA can bind to
    & RNA sequence
    & Result online
    & RBPs name
    & Web server
    & Predict RNA binding only to the 68 pre-defined RBPs in the library
    &\ding{72}\ding{72}\ding{72}\ding{72}\ding{73}
    \\\hline
    
    GraphBind ~\cite{xia2021graphbind} & \url{http://www.csbio.sjtu.edu.cn/bioinf/GraphBind/} & It provides the prediction for structure-based nucleic acid- and small ligand-binding residues
    & PDB
    & CSV 
    & Binding residues, 3D binding interface
    & Web server
    & It only supports processing one sequence at a time. It only predicts specific ligand types that it was trained on
    &\ding{72}\ding{72}\ding{72}\ding{72}\ding{73} 
    \\\hline
    
    iDRNA-ITF ~\cite{wang2022idrna} & \url{http://bliulab.net/iDRNA-ITF} & It provides the prediction for DNA- or RNA-binding residues on proteins 
    & FASTA (protein)
    & Result online, TXT
    & Binding residues, binding probability/labels
    & Web server
    & Input sequence restrictions, ligand type constraints
    &\ding{72}\ding{72}\ding{72}\ding{73}\ding{73}  
    \\\hline
    
    Pprint2 ~\cite{patiyal2023prediction}  & \url{https://webs.iiitd.edu.in/raghava/pprint2} & It provides the prediction for RNA interacting protein residues
    & FASTA (protein)
    & Result online, CSV
    & Binding propensity score, binary label
    & Web server, standalone version
    & Sequence length restrictions, Web server concurrency constraints, limited accuracy
    &\ding{72}\ding{72}\ding{72}\ding{72}\ding{73}
    \\\hline
    
    NABind ~\cite{jiang2023structure} & \url{http://liulab.hzau.edu.cn/NABind/} & It provides the prediction for DNA or RNA interacting protein residues 
    & PDB
    & Result online, GIF, JPG, PNG, POV, 3D
    & Binding residues, binding propensity score, binary label
    & Web server, standalone version, docker image 
    & Exclusively for nucleic acid binding, restrictions on sequence length and batch size,  non-trivial local deployment
    &\ding{72}\ding{72}\ding{72}\ding{73}\ding{73}
    \\\hline
    
    RNAincoder ~\cite{wang2023rnaincoder}  & \url{https://idrblab.org/rnaincoder/} & It provides the encoding results of RNA for RPIP
    & FASTA (RNA) 
    & CSV, TSV
    & RNA sequence encoding
    & Web server
    & Sequence length limitations, batch submission quantity restrictions
    &\ding{72}\ding{72}\ding{72}\ding{72}\ding{73}
    \\\hline
\end{longtable}

Many studies have developed their models into software or websites for other researchers to use. 
These RPIP tools are listed in Table~\ref{tab:tools_analysis}.

\subsection{Tool functionality}

We analyzed the tools in Table~\ref{tab:tools_analysis} according to their purpose or function.

\textit{Interaction prediction:}
Users can upload protein structures in the PDB file format for predicting the binding preferences of the RNA components (phosphate and ribose), at any location on a given protein surface.
These tools can be used to perform RPIP.

\textit{Prediction of binding sites:}
Users can input protein names or sequences, set parameters and thresholds, and obtain the starting and ending positions, sequences, and probability values of potential binding sites (with probabilities greater than the set threshold).

\textit{RBP prediction:}
Users can input protein sequences to determine whether or not the protein is an RBP.

\begin{tcolorbox}[colback=white,sharp corners,boxrule=1pt]
    \textit{\textbf{Summary of answers to RQ7:}}
    \begin{itemize}
        \item[(1)] 
        RPIP tools cover many different prediction needs, and have many different potential applications.
        
        \item[(2)]
        Tool performance is constrained by data scale, species coverage, and dynamic modeling capabilities.
        Usually, the data comes from the tool developers.

        \item[(3)]
        The interpretability of tools is generally poor.
    \end{itemize}
\end{tcolorbox}

\section{Answer to RQ8: Remaining Challenges and Future Work for RPIP
\label{SEC:rq8}}

This section discusses the remaining challenges and potential future work for RPIP.

\subsection{Key challenges}
\textbf{\textit{Challenge 1: Lack of Unified Public Databases.}}
Databases related to RNAs and proteins may have inconsistent query methods, making information cross-validation difficult. 
Due to insufficient biological knowledge, interdisciplinary researchers may not easily be able to efficiently obtain cross-database information (such as for databases like RPI369 and RPI488, which require cross-retrieval).

\textbf{\textit{Challenge 2: Deficiencies in Dataset Balancing Strategies.}}
RPI datasets often suffer from a lack of negative samples, and may have inconsistent criteria for constructing them:
RPI369 and similar datasets, for example, define interactions based on atomic distance; 
NPInter relies on experimental detection; and
RBP68 uses binding residue distance. 
This inconsistency can lead to biases in model training.

\textbf{\textit{Challenge 3: Diversity of RNA or Protein Features and Encoding.}}
Inconsistent RNA or protein-sequence lengths necessitate encoding methods that capture nucleotide or amino acid dependencies.
Molecular structural effects render traditional distance metrics (e.g., Euclidean) inadequate:
$K$-mer encoding ignores positional information and sequence similarity.

\textbf{\textit{Challenge 4: Limitations of DL Models.}}
The high cost of biological experiments can restrict high-quality RPI data availability, hampering model generalization.
Dataset imbalances cause prediction bias toward negative samples, reducing positive sample accuracy.
The black-box nature of DL models impedes interpretability and cross-dataset transferability.

\textbf{\textit{Challenge 5: Absence of Multi-Dataset Validation Frameworks.}}
RNA and protein datasets are highly complex. 
Current validation methods, like cross - validation, cannot fully assess model performance. 
This is especially true for varying sequence lengths. 
It also applies to situations with larger structural complexity.

\textbf{\textit{Challenge 6: Difficulty in Model Deployment.}}
Current RPIP tools operate as black boxes, hiding details, and making interpretability difficult. 
Their datasets are primarily developer-curated, resulting in poor generalizability and a lack of standardized deployment frameworks.

\subsection{Approaches/tools to address RPIP challenges}

Research has identified core challenges for RPIP at the data, model, validation, and application levels.
There have been attempts to address some challenges (such as few-shot learning and cross-species migration) with tools like prototype networks and transfer learning models. 
However, issues like unified database construction and dataset balancing strategies still lack standardized tools, which need further exploration in future research. 

\begin{table}[!t]
    \caption{Current challenges and potential solutions/tools in RPIP research.}
    \label{tab:Challenges_solutions}
    \centering
    \begin{threeparttable}
        \scriptsize
        \renewcommand{\arraystretch}{1.4} 
        \setlength{\tabcolsep}{1mm}{
        \begin{tabular}{m{1.5cm}|m{1.8cm}|m{2.8cm}|m{6.5cm}}
            \hline
            \textbf{Challenges} & \textbf{Core issue} & \textbf{Potential solutions / tools}& \textbf{Brief description} \\\hline
            Challenge 1  &  Fragmented data resources & BioGRID ~\cite{oughtred2019biogrid}
            & A highly-curated PPI database with multi-species coverage
            \\\hline
            Challenge 2   & Lack of negative sample standards & Degree distribution balanced sampling ~\cite{li2025negative} 
            & Aligning the node degree distribution of negative samples with positive samples according to the scale-free property of biomolecular networks
            \\\hline
            Challenge 3   & Feature encoding limitations & Multimodal language models, ESM~\cite{lin2023evolutionary}, RNAFM~\cite{chen2022interpretable} 
            & Pretrained Language Models
            The pretraining-finetuning paradigm not only captures contextual information but also mitigates data scarcity challenges
            \\\hline
            Challenge 4   & Model generalization and interpretability & Multimodal, transfer learning 
            & Integrating multimodal and transfer learning deep learning architectures can, to some extent, mitigate the model's generalization limitations
            \\\hline
            Challenge 5  & Non-unified validation protocols
            & BioGRID, STRING 
            & Augmenting BioGRID's cross-species RPIs using STRING's functional enrichment tools
            \\\hline
            Challenge 6   & Immature tool ecosystem 
            & Captum\cite{kokhlikyan2020captum}, RNAFormer\cite{franke2024rnaformer}, MEME\cite{bailey2009meme} 
            & Captum utilizes attention mechanism visualization to discover critical nucleotides
            Captum Insights offers a web-based interactive interface that enables researchers to intuitively debug models through methods such as heatmaps and feature importance rankings
            RNAFormer maps prediction results onto RNA secondary structures, and MEME extracts RNA motifs identified by the model for matching against known databases
            This is not a task that can be accomplished by a single tool, but requires an ecosystem engineering approach
            \\\hline
        \end{tabular}  
        }  
    \end{threeparttable}
\end{table}

Some mature technologies and tools may represent potential solutions for the remaining challenges for RPIP.
For fragmented data resources, the highly manually-curated \textit{Biological General Repository for Interaction Datasets} (BioGRID) database~\cite{oughtred2019biogrid} can be used to integrate multi-species interaction data.
A degree-distribution-balanced sampling strategy~\cite{li2025negative}, based on the scale-free property of biomolecular networks, may be able to address the lack of negative sample standards.
Feature-encoding limitations may be addressed by multimodal pre-trained language models, such as \textit{Evolutionary Scale Modeling} (ESM)~\cite{lin2023evolutionary} and \textit{RNA Foundation Model} (RNAFM)~\cite{chen2022interpretable}:
The pre-training-fine-tuning paradigm can be used to capture the deep-level semantic information of sequences.
For insufficient model generalization and interpretability, it may be necessary to construct a multimodal fusion and transfer learning architecture to enhance the cross-scenario adaptability.
For the non-unified validation protocols, the functional enrichment tools of BioGRID and STRING can be used for collaborative validation.
For the immature tool ecosystem, a tool chain consisting of Captum~\cite{kokhlikyan2020captum} (for visualizing key nucleotides), \textit{RNA Fold Oriented Representation Model with Transformer} (RNAFormer) \cite{franke2024rnaformer} (for mapping RNA secondary structures), and MEME~\cite{bailey2009meme} (for motif matching) could form an interpretable framework.
Table~\ref{tab:Challenges_solutions} lists the current challenges and potential corresponding solutions or tools.

\subsection{Future work}

\textbf{\textit{Improve data quality:}}
It is necessary to develop new data-acquisition and annotation techniques to expand the training datasets and improve data quality. 
If the problems of data scarcity and imbalance can be alleviated, high-quality datasets will greatly increase the accuracy of the models.

\textbf{\textit{Feature-representation exploration:}}
It is necessary to introduce more advanced feature-extraction and representation methods to comprehensively capture the features of RNAs and proteins. 
Combining multiple types of RNA and protein information, and using multimodality techniques, will also improve the accuracy and robustness of RPIP.
The features and coding methods for RNA and proteins are diverse:
Exploring optimal combinations of features, and developing new encoding methods will be meaningful research directions.

\textbf{\textit{Few-shot learning:}}
Although DL has achieved good results in RPIP, it requires a large amount of data.
Few-shot learning trains models using limited datasets, and enables effective prediction of new categories. 
This is particularly important for RPIP, as the data is often scarce and costly to obtain. 
Through few-shot learning, the model can learn effective feature representations using only a small number of samples, and achieve good RPIP performance. 
The prototypical network~\cite{zhao2022predicting} is used to predict the RPI, which indicates that few-shot learning is feasible for RPIP.
In limited-data scenarios, hybrid data-augmentation methods~\cite{osmani2025deflection} combining intelligently optimized features with physics-informed modeling~\cite{ouadi2024optimizing} have demonstrated significantly enhanced learning capacity.

\textbf{\textit{Transfer learning:}}
Transfer learning can be used to solve the cross-species problem~\cite{zhang2022prerbp}. 
Deep transfer learning uses pre-trained models to extract effective feature representations from RNAs and proteins. It can then transfer the learned parameters or feature representations to other datasets for RPIP.
Species-specific RBP-recognition methods based on transfer learning have been introduced~\cite{zhang2022prerbp,peng2022rbp}.
Zhang et al.~\cite{zhang2024cbil} first trained their model on a large-scale, diversified dataset encompassing plants, animals, and other biological taxa. 
Following pretraining, transfer learning was implemented on a specialized viral protein-human LncRNA interaction dataset. 
The results confirmed the model's efficacy in cross-species prediction tasks.
In RPIP, deep transfer learning can help models to adapt to different types of RNAs and proteins, and to predict interactions under various conditions.

\textbf{\textit{Optimizing the model validation methods:}}
During verification, it may be found that the model performs poorly in certain specific situations, such as with varied sequence lengths and increasing structural complexity. 
DL models often have low interpretability because of their complex network structure and nonlinear mapping relationships. 
More deeply exploring the DL-model mechanisms may help improve their interpretability.

\textbf{\textit{Tool development trends:}}
RPIP tools can train more generalized models by collecting or integrating new experimental data.  
To meet the needs of biological researchers, RPIP tools should place greater emphasis on model interpretability.
With better interpretability, the biological mechanisms behind model predictions can be revealed, which will better support experimental research. 
With further DL development, we believe that this field will achieve more fruitful results.

\begin{tcolorbox}[colback=white,sharp corners,boxrule=1pt]
    \textit{\textbf{Summary of answers to RQ8:}}
    \begin{itemize}
        \item[(1)] 
        We have identified six current challenges for DL-based RPIP that will require further investigation.
        Current solutions focus on partial optimizations of data, models, and validation:
        They need to overcome the dual challenges of ``generalization-interpretability''.
        
        \item[(2)] 
        Future research should investigate technical integration and innovation, strengthening the connections between DL models and biological experiments.
        Further research in DL-based RPIP is needed in terms of data quality, RNA/protein features and encoding, more effective model optimization and validation, and better tool development.      
    \end{itemize}
\end{tcolorbox}

\section{Comparison with Previous Review Studies}
\label{SEC:Related_Survey}

\begin{table}
    \caption{Comparison between our work and previous review studies. 
    (\textbf{NR} means not reported; 
    \ding{52} and \ding{56} indicate whether or not the corresponding aspect has been summarized; and 
    \textbf{RQ1} to \textbf{RQ8} refer to our eight research questions, from Section~\ref{SEC:Methodology}.)}
    \scriptsize
    \centering
    \label{tab:previous review}
    \setlength{\tabcolsep}{1mm}{
    \begin{tabular}{m{1.4cm}|c|m{0.6cm}|c|m{2cm}|m{0.6cm}|m{1.1cm}|c@{\hspace{2pt}}c@{\hspace{2pt}}c@{\hspace{2pt}}c@{\hspace{2pt}}c@{\hspace{2pt}}c@{\hspace{2pt}}c@{\hspace{2pt}}c}\hline
        \textbf{Reference}&\textbf{Year}& \textbf{Time span}&\textbf{Studies}&\textbf{RPI dimensions}&\textbf{DL only}&\textbf{Covered DL methods}&\textbf{RQ1}&\textbf{RQ2}&\textbf{RQ3}&\textbf{RQ4}&\textbf{RQ5}&\textbf{RQ6}&\textbf{RQ7}&\textbf{RQ8}
        \\\hline                       
        Pan et al. \cite{pan2019recent} &  2019 & NR & NR  & Prediction of RPI pairs and RBP-binding sites on RNAs &\ding{56} &CNN  RNN &\ding{56} &\ding{52} &\ding{52} &\ding{52} &\ding{56} &\ding{56} &\ding{52} &\ding{52} 
        \\\hline
        Moore et al. \cite{moore2019computational} & 2019 & NR & NR & Prediction of RPI pairs and RBP-binding motifs & \ding{56} & CNN  RNN  DNN &\ding{56} &\ding{56} &\ding{56} &\ding{52} &\ding{56} &\ding{56}&\ding{56} &\ding{56} 
        \\\hline
        Zhang et al. \cite{zhang2019long} & 2019 & NR & NR & Prediction of RPI pairs & \ding{56} & GNN &\ding{56} &\ding{52} &\ding{56} &\ding{52} &\ding{56} &\ding{56}&\ding{56} &\ding{56} 
        \\\hline
        Yan et al. \cite{yan2020review} & 2020 & NR & NR & Prediction of proteins binding sites on RNAs & \ding{56} & CNN  
        RNN  DBN  GCN  GAN  &\ding{56} &\ding{52} &\ding{52} &\ding{52} &\ding{56} &\ding{52}&\ding{56} &\ding{52} 
        \\\hline
        Xia et al. \cite{xia2021recent} &  2021 & NR & NR  & Prediction of binding sites or motifs, RNA-protein docking methodology, RPI mechanism analysis &\ding{56} &CNN  RNN &\ding{56} &\ding{56} &\ding{52} &\ding{52} &\ding{56} &\ding{56} &\ding{52} &\ding{56} 
        \\\hline
        Philip et al. \cite{philip2021survey} & 2021 & 2016-2021 & NR & Prediction of RPI pairs & \ding{56} & CNN  RNN &\ding{56} &\ding{52} &\ding{56} &\ding{56} &\ding{52} &\ding{56}&\ding{56} &\ding{56} 
        \\\hline
        Zhong et al. \cite{zhong2021recent} & 2021 & 2016-2020 & NR & Prediction of RPI pairs & \ding{56} & SAN &\ding{56} &\ding{52} &\ding{52} &\ding{52} &\ding{52} &\ding{56}&\ding{56} &\ding{52} 
        \\\hline
        Wei et al. \cite{wei2022protein} & 2022 & 2004-2021 & NR & Prediction of binding site and binding preference & \ding{56} & CNN  RNN  GNN &\ding{56} &\ding{52} &\ding{52} &\ding{52} &\ding{52} &\ding{52}&\ding{56} &\ding{52} 
        \\\hline
        Shaath et al. \cite{shaath2022long} & 2022 & NR & NR & Prediction of RPI pairs & \ding{56} & CNN  RNN  GNN &\ding{56} &\ding{56} &\ding{56} &\ding{52} &\ding{56} &\ding{56}&\ding{56} &\ding{56} 
        \\\hline
        Cui et al. \cite{cui2022protein} & 2022 & NR & NR & Prediction of protein–DNA/RNA interaction pairs & \ding{56} & CNN  RNN &\ding{56} &\ding{56} &\ding{56} &\ding{52} &\ding{56} &\ding{56}&\ding{52} &\ding{56} 
        \\\hline
        Lasantha et al. \cite{lasantha2023deep} & 2023 & 2014-2023 & NR & Only cicRNA-protein interactions & \ding{56} & CNN  RNN  &\ding{56} &\ding{56} &\ding{56} &\ding{52} &\ding{56} &\ding{56}&\ding{56} &\ding{52} 
        \\\hline
        \textbf{This study} & - & 2014-2023 & 179 & Prediction of RPI pirs, prediction of binding sites or motifs, prediction of the RBP, etc. & \ding{52} & CNN  RNN  GAN  DBN  DAE  MLP  GNN Transformer &\ding{52} &\ding{52} &\ding{52} &\ding{52} &\ding{52} &\ding{52}&\ding{52} &\ding{52} 
        \\\hline   
    \end{tabular}}
\end{table}

\subsection{Analysis of previous review studies}
In recent years, significant progress has been made in research on RPIP, and the number of review papers in related fields has also grown. 
Previous review papers have compiled and examined research achievements from various perspectives. 
Table~\ref{tab:previous review} provides a detailed comparison of the differences between our study and previous reviews, across multiple dimensions.
We have systematically reviewed the applications of DL methods in RPIP, and conducted a comprehensive examination of each stage of the prediction workflow. 
As shown in Table~\ref{tab:previous review}, previous studies differ from this research in terms of research scope, methodological focus, and the included stages of the prediction process. 
Other reviews typically provide comprehensive overviews, but differ from those of this study in terms of focal points. 
For example, Pan et al.~\cite{pan2019recent} focused on two types of RPIP tasks: 
predicting whether or not RNA and proteins bind, and identifying RBP-binding sites. 
Their review also covers ML methods, such as \textit{Support Vector Machine} and \textit{Random Forest,} in RPIP research. 
However, their section on DL methods only includes CNN and RNN, omitting models that have appeared since 2020.
Regarding the introduction of RPI-prediction methods, Xia's~\cite{xia2021recent} work leans more toward explaining the biological mechanisms of RNA-protein interactions.
Philip's review~\cite{philip2021survey} of predicting RNA-protein interactions primarily focuses on ML methods.
Yan's~\cite{yan2020review} research emphasizes the application of ML and DL specifically in RNA-protein binding-site prediction.
Shaath~\cite{shaath2022long} evaluated prediction methods for binding type intrinsically disordered regions in interactions with various ligands (peptides, proteins, RNA, DNA, lipids).
Cui's~\cite{cui2022protein} emphasizes the differences between RNA-protein interactions and DNA-protein interactions.

To the best of our knowledge, this is the most comprehensive review examining the entire process of RPI prediction using DL. 
As shown in Table~\ref{tab:previous review}, other studies not only do not exclusively address DL methods in RPIP research, but also provide very limited coverage of DL approaches, particularly lacking discussions on recent and popular DL techniques. 
None of the other studies fully address all of our research questions (RQ1 to RQ8).
Our study also includes a comprehensive analysis of all aspects of the entire prediction workflow, offering a broader perspective.
We have comprehensively reviewed cutting-edge DL technologies, highlighting their practical application value, integrating comprehensive feature-encoding methods and RPIP tools, as well as discussing their limitations. 
This review provides practical reference resources for researchers and practitioners in terms of dataset preprocessing, data-feature encoding methods, adaptation of DL models to tasks, and tool selection.

\subsection{Identification of research gaps}
Through a comparison of previous reviews (Table ~\ref{tab:previous review}), we have identified the following research gaps:

\subsubsection{Limited coverage/scope}

Previous reviews (e.g., Pan et al. 2019, Xia 2021) only focused on partial tasks (such as binding-site prediction) and did not systematically cover the entire DL-based RPIP process (feature encoding $\rightarrow$ model construction $\rightarrow$ evaluation $\rightarrow$ applications $\rightarrow$ tools $\rightarrow$ challenges).

\subsubsection{Incomplete model coverage}
Most reviews only analyzed traditional models like CNNs and RNNs (covering $\leq 5$ types), and lacked discussion on emerging DL models such as Transformers, GNNs, and GANs (e.g., Philip 2021 does not address GNNs).

\subsubsection{Missing key components}
Previous review studies failed to adequately address eight critical RQs, especially for bibliometric analysis (RQ1), dataset-bias assessment (RQ2), cross-species validation (RQ5), tool usability (RQ7), and current challenges analysis (RQ8).

\subsection{Clarification of research objectives}

To address the research gaps, this review has aimed to:
(1) establish the most comprehensive framework for DL-based RPIP surveys;
(2) provide a bibliometric analysis of research trends and scholarly connections within the field;
(3) systematically perform bias assessment on commonly used datasets;
(4) examine feature-encoding strategies, diverse DL architectures, and hybrid methodologies;
(5) critically analyze model limitations and application scenarios;
(6) assess the cross-species generalization capabilities of validation models; and
(7) evaluate practical applications and open-source tools
---
detailing their scalability, constraints, and black-box prediction; and
(8) identify remaining and emerging challenges, and examine potential solutions.

\subsection{Contributions and novelty of the study}
\label{SEC:Contributions and Novelties}

\subsubsection{Entire-process analysis framework}

This study is, we believe, the most comprehensive systematic survey on the entire process of DL-based RPIP.
It covers the complete workflow from dataset preprocessing and feature encoding to model construction, result evaluation, and application scenarios. 
Our eight research questions address dimensions such as publication trends, data characteristics, model architectures, and tool applications.

\subsubsection{Bibliometric contributions of DL-based RPIP research}

We have systematically quantified DL-based RPIP research, identifying sustained linear growth trends.
We have provided a comprehensive review of the research thematic network through authors' keywords co-occurrence analysis.
This study has identified the main scholars and collaboration networks within DL-based RPIP, uncovering interdisciplinary collaboration patterns between theoretical development and engineering applications.
The presented literature co-citation graphs have also mapped out research trajectories and highlighted fundamental publications.

\subsubsection{Dataset and database collation}

Database integration: Core databases (such as NONCODE, CircBase, Uniprot, and NPInter) were examined, and their data types and applicable scenarios analyzed.

Dataset construction: Methods for constructing high-frequency datasets such as RPI488 and RBP31 have been summarized (e.g., division of positive and negative samples based on atomic distance thresholds), distinguishing between structure-driven types (e.g., RPI488/RPI1807 based on an atomic distance threshold of 3.4~\AA) and experimentally validated types (e.g., NPInter). 
A unified methodology for constructing negative samples was summarized, and we discussed negative sample-generation strategies (e.g., random sampling, sequence shuffling) and dataset imbalance issues.

\subsubsection{Novel feature-encoding strategies}

This paper has systematically explored encoding methods, such as one-hot encoding, $k$-mer frequency, and word embedding (e.g., RNA2vec and Pro2vec), and discussed enhancing feature representation by combining physicochemical properties (e.g., hydrophobicity and charge) and evolutionary information (e.g., PSSM matrix). 
Outputs from tools such as RNAfold (secondary-structure prediction) and AlphaFold (protein-structure prediction) have been discussed.
We have also examined structural information being converted into character sequences for processing using sequence-encoding methods, with an emphasis on the joint use of sequence and structural features.

\subsubsection{Analysis of multiple DL frameworks}

The adaptability and limitations of various DL models for RPIP tasks have, for the first time, been discussed. 
The motif-detection capability of CNN is a unique advantage for RPIP. 
GNN was found to be suitable for sparse data and molecular graph analysis. 
We have also provided some discussion about Transformers capturing global dependencies through self-attention mechanisms, and how pre-trained models (such as DNABERT) can be migrated to RPIP tasks. 
Hybrid model architectures that fuse CNN$+$LSTM$+$attention mechanism help improve the comprehensiveness of feature learning.

\subsubsection{Cross-species validation-method analysis}

This survey has also addressed cross-database, cross-species (e.g., human and plant datasets), and in-vivo/in-vitro experimental validation. 
We highlighted that cross-species validation is underused due to data scarcity and species-specific feature differences
---
we also noted that it is critical for model generalizability. 
The diversity of RNA-protein interaction mechanisms across species necessitates integrating domain knowledge to optimize model adaptability.

\subsubsection{Tool and application integration}

This survey has also summarized online tools and discussed tools such as \textit{CircRNA-RBP Web Server} (CRWS) (circular RNA binding site prediction), \textit{Deep Contextual Learning for Identifying Protein-RNA Interaction Profiles} (DeepCLIP) (binding motif analysis), and \textit{Protein-RNA Interaction by Structure-informed Modeling using deep neural NETwork} (PrismNet). 
These tools were classified into three categories by function: 
interaction prediction (e.g., \textit{Interaction Pattern Miner} (IPMiner)); 
binding-site detection (e.g., \textit{Protein Surface Topography for Protein-RNA Binding Sites Prediction} (PST-PRNA)); and 
RBP identification (e.g., \textit{Predictor of DNA-Binding Proteins and RNA-Binding Protein-Multilayer Perceptron} (PredDRBP-MLP)).
Applications have covered areas such as cancer-related RPI identification (e.g., for bladder cancer and \textit{Severe Acute Respiratory Syndrome Coronavirus 2} protein-binding sites), RNA/protein function annotation, and cross-species interaction prediction, highlighting the practical value of DL in biomedicine.

\subsection{Limitations of the study and future research directions}

Some possible \textit{limitations} of this survey are as follows.
(1) Restricted database selection: 
Only six literature repositories were screened, ignoring preprint platforms (e.g., bioRxiv) and other possible databases.
(2) Time span constraint: 
This study was limited to 2014–2023. 
Although reasonable, this inevitably means that some early studies have been omitted.
(3) Language filtering: 
Only English literature was included, potentially missing important achievements in other languages.
(4) Model-scope limitation: 
This paper focuses on DL in RPIP. 
Other approaches, including broader ML-based RPIP, while also valuable, were not included in the analyses.

Possible \textit{future directions} for us include that we will examine literature from before 2014 and after 2023, from platforms like bioRxiv and Nature.
We will further analyze non-English literature using translated materials and cross-lingual learning techniques.
Finally, we will also study hot technologies and related models, integrating key breakthroughs for comparison or fusion with DL.

\section{Conclusion
\label{SEC:Conclusion}}

In this survey, we investigated DL-based RPIP studies published between 2014 and 2023. 
We selected six major databases for the literature search.
Finally, 179 research papers were selected for analysis.
Eight RQs were used to guide the survey.
We classified and statistically analyzed the evolution and distribution of the DL-based RPIP literature. 
The survey covers the entire process, including datasets, RNA and protein features-encoding methods, DL models, and model evaluation metrics.
We also studied DL-based RPIP application areas and tools, highlighting that it has been applied in multiple fields, and that there have been several DL-based RPIP services and tools made available. 
We have also discussed some current challenges that require further investigation, and identified some directions for related future work. 
We believe that this article can provide strong guidance for the future development of DL-based RPIP.

\section*{Acknowledgements}

We would like to thank the anonymous reviewers for their many constructive comments.
This work was partially supported by the Science and Technology Development Fund of Macau, Macau SAR, under Grant Nos. 0021/2023/RIA1 and 0035/2023/ITP1.

\bibliographystyle{elsarticle-num}
\bibliography{RPIP}

\end{document}